\documentclass[12pt, a4]{article}
\usepackage{amsfonts}
\usepackage{graphicx}
\usepackage{amsmath}
\usepackage{amssymb}

\begin{document}

\title{\textbf{Kinetic equation for a soliton gas and its hydrodynamic
reductions}}
\author{G.A.~El$^1$, \ A.M.~Kamchatnov$^2$,\ M.V.~Pavlov$^3$ and S.A.~Zykov$%
^4$ \\
\\
$^1$ Department of Mathematical Sciences, Loughborough University, UK\\
$^2$Institute of Spectroscopy, Russian Academy of Sciences,\\
Troitsk, Moscow Region, Russia \\
$^3$ Lebedev Physical Institute, Russian Academy of Sciences, Moscow \\
$^4$ SISSA, Trieste, Italy, and\\
Institute of Metal Physics, Urals Division of Russian Academy of Sciences, \\
Ekaterinburg, Russia}
\date{}
\maketitle

\begin{abstract}
We introduce and study a new class of kinetic equations, which arise in the
description of nonequilibrium
macroscopic dynamics of soliton gases with elastic collisions  between solitons. These
equations represent nonlinear integro-differential systems and have a novel
structure, which we investigate by studying in detail the class of $N$-component
`cold-gas' hydrodynamic reductions. We prove that these
reductions represent integrable linearly degenerate hydrodynamic type
systems for arbitrary $N$ which is a strong evidence in favour of integrability of
the full kinetic equation. We derive compact explicit representations for the
Riemann invariants and characteristic velocities of the hydrodynamic
reductions in terms of the `cold-gas' component densities and construct a
number of exact solutions having special properties (quasi-periodic,
self-similar). Hydrodynamic symmetries are then derived and investigated.
The obtained results shed the light on the structure of a continuum limit for a large class of integrable systems
of hydrodynamic type and are also relevant to the description of turbulent motion in conservative compressible flows.

\end{abstract}


\section{Introduction and summary of  results}

  The possibility of modelling  certain types of turbulent motion with the aid of the equations for  weak limits of highly oscillatory dispersive compressible flows (the so-called Whitham modulation equations \cite{wh74}, \cite{ffm79}, \cite{laxlev83}) was pointed out by P.D. Lax in \cite{lax91}. While this ``deterministic analogue of turbulence''  has obvious limitations to its possible applications to the description of hydrodynamic (incompressible) turbulent flows, it opens  a whole new perspective for constructing the statistical description
of purely conservative wave regimes in integrable dispersive systems by assigning appropriate probabilistic measures to the wave sequences, so that their weak limits could then be regarded as {\it ensemble averages}.
Such an unconventional union of integrability and
stochasticity has a natural physical motivation: nonlinear dispersive waves,
while often being successfully modelled by integrable systems, could
demonstrate  a very complex behaviour calling for a statistical description
characteristic of the classical turbulence theories.
Recently, a closely related   programme for
the construction of the theory of {\it wave turbulence} in the frameworks of
integrable systems has been put forward by V.E. Zakharov  \cite{zakh09}.

One of the important problems arising in this connection is the description
of ``soliton gases'' --- random distributions of solitons which can be
mathematically defined in terms of generalised reflectionless potentials
with shift invariant probability measure on them (see e.g. \cite{kot08}).
Due to isospectrality of the ``primitive'' microscopic evolution, the
macroscopic dynamics of a homogeneous soliton gas is trivial (for the
so-called `strongly integrable' systems, such as the Korteweg -- de Vries
(KdV), nonlinear Schr\"odinger (NLS) or Kadomtsev-Petviashvili (KP-II)
equations --- see \cite{zakh09}), namely, all the statistical
characteristics can be specified arbitrarily at the initial moment and
remain unchanged in time. However, if the soliton gas is spatially
inhomogeneous, i.e. if the probability distribution function depends on the
space coordinate, then nontrivial macroscopic dynamics can occur due to phase
shifts of individual solitons  in their collisions with each other. An
approximate kinetic equation describing spatial evolution of the soliton
distribution function in a rarefied gas of the KdV solitons, when these
phase shifts can be taken into account explicitly, was derived by Zakharov
back in 1971 \cite{zakh71}.

Generalization of Zakharov's kinetic equation to the case of a soliton gas
of finite density has been made possible rather recently \cite{el03} and
required consideration of the continuum \textit{\ thermodynamic} limit of
the Whitham modulation equations associated with finite-gap potentials.
In the thermodynamic limit, the nonlinear interacting wave modes
transform into randomly distributed localised states (solitons) and the
modulation system assumes the form of a nonlinear kinetic equation. This new
kinetic equation was extended, using physical reasoning, in \cite{elkam05}
to other integrable systems with two-particle elastic interactions of
solitons (i.e. when multi-particle effects are absent).

The kinetic equation for solitons in general form represents a nonlinear
integro-differential system
\begin{equation}
\begin{split}
& f_{t}+ (sf)_{x} = 0\,, \\
& s(\eta ) =S(\eta )+\frac{1}{\eta }\int\limits_{0}^{\infty }G(\eta ,\mu
)f(\mu )[s(\mu )-s(\eta )]d\mu \,.
\end{split}
\label{kin1}
\end{equation}%
Here $f(\eta )\equiv f(\eta ,x,t)$ is the distribution function and $s(\eta
)\equiv s(\eta ,x,t)$ is the associated transport velocity. The (given)
functions $S(\eta )$ and $G(\eta ,\mu )$ do not depend on $x$ and $t$. The
function $G(\eta ,\mu )$ is assumed to be symmetric, i.e. $G(\eta ,\mu ) =
G(\mu, \eta)$. The choice
\begin{equation}
S(\eta )=4\eta ^{2}\,,\qquad G(\eta ,\mu )=\log \left\vert \frac{\eta -\mu }{%
\eta +\mu }\right\vert  \label{kdv}
\end{equation}%
corresponds to the KdV soliton gas \cite{el03}, where the KdV equation is taken in the canonical form
\begin{equation}  \label{kdveq}
\phi _{t}-6\phi \phi _{x}+\phi _{xxx}=0 \, .
\end{equation}
In the KdV context, $\eta \ge 0 $ is a real-valued spectral parameter (to be
precise, before the passage to the continuum limit one has $%
\lambda_k=-\eta_k^2$, where $\lambda_k$, $k=1, \dots, N$ are the discrete
eigenvalues of the Schr\"odinger operator), thus the function $f(\eta ,x,t)$
is the distribution function of solitons over spectrum so that $\kappa
=\int_{0}^{\infty }f(\eta )d\eta =\mathcal{O}(1)$ is the spatial density of
solitons. If $\kappa \ll 1$, the first order approximation of (\ref{kin1}),
(\ref{kdv}) yields Zakharov's kinetic equation for a dilute gas of KdV
solitons \cite{zakh71}  (see equations (\ref{kin11}), (\ref{s1z}) below).

The quantity $S(\eta )$ in (\ref{kin1}) has a natural meaning of the
velocity of an isolated (free) soliton with the spectral parameter $\eta $
and the function $\frac{1}{\eta}G(\eta ,\mu )$ is the expression for a phase
shift of this soliton occurring after its collision with another soliton
having the spectral parameter $\mu <\eta $. Then $s(\eta, x,t)$ acquires the
meaning of the self-consistently defined mean local velocity of solitons
with the spectral parameter close to $\eta$ (see \cite{elkam05}).

Theory of nonlocal kinetic equations of the form (\ref{kin1}) is not
developed yet. Possible approaches to their treatment were discussed in \cite%
{belokolos05} in connection with special classes of exact solutions for the
Boltzmann kinetic equation for Maxwellian particles. The derivation of (\ref%
{kin1}), (\ref{kdv}) as a certain (albeit singular) limit of the integrable
KdV-Whitham system suggests that this new kinetic equation is also an
integrable system, at least for special choices of functions $G(\eta ,\mu )$%
. A natural question arising in this connection is: what is the exact
meaning of integrability for the equations of the type (\ref{kin1})?

Integrability of kinetic equations has been the object of intensive studies
in recent decades. For instance, integrability of the collisionless
Boltzmann equation (which is sometimes called the Vlasov equation) can be
defined in terms of two other closely connected (even equivalent in some
sense) objects: the Benney hydrodynamic chain \cite{benney73}, \cite{zakh81}%
, \cite{gib81} and the dispersionless limit of the Kadomtsev--Petviashvili
equation (\cite{kod88a, kod88b}, \cite{gibkod94}. It turns out that all these three different
nonlinear partial differential equations possess the same infinite set of $N$%
-component hydrodynamic reductions parameterised by $N$ arbitrary functions
of a single variable \cite{gibts96, gibts99} (we note that the solutions to
these $N$-component reductions are parameterised, in their turn, by another $%
N $ arbitrary functions of a single variable). This property was used in
\cite{ferkhus04a, ferkhus04b}(see also \cite{zakh94}, \cite{fermar07}, \cite%
{gibrai07}, \cite{odpavsok08}) when introducing the integrability criterion
for a wide class of kinetic equations, corresponding hydrodynamic chains and
2+1 quasilinear equations. Moreover, it was proved in \cite{pav07} that the
existence of at least one $N $-component hydrodynamic reduction written in
the so-called \textit{symmetric} form is sufficient for integrability in the
sense of \cite{ferkhus04a}. Another possible approach to analyse an
integrable kinetic equation is to use the fact that it possesses infinitely
many particular solutions determined by the corresponding hydrodynamic
reductions (see \cite{odpavsok08} for details).

The distinctive feature of the kinetic equation (\ref{kin1}) is its nonlocal
structure, which represents an obstacle to the direct application to it of
the approaches developed in \cite{pav07} and \cite{odpavsok08}. For
instance, the possibility of an explicit construction of symmetric
hydrodynamic reductions, and even the existence of such reductions for (\ref%
{kin1}), are open questions at the moment. In this paper, we study a
particular, yet probably the most important from the viewpoint of capturing
the essential properties of the full equation, family of the `cold-gas' $N$
-component hydrodynamic reductions of (\ref{kin1}) obtained via the
delta-function ansatz for the distribution function $f(\eta, x,t) =
\sum_{i=1}^{N}f^{i}(x,t)\delta (\eta -\eta _{i})$, where $\eta_N >
\eta_{N-1}> \dots > \eta_1>0$ are arbitrary numbers.  Then the velocity distribution $s(\eta, x,t)$ over the `spectrum'
becomes a discrete set of functions $\{s^i(x,t): \ s^i=s(\eta_i,x,t), \ i=1, \dots, N \}$ and the sought reductions family
assumes the form of a system of hydrodynamic conservation laws
\begin{equation}  \label{01}
u_{t}^{i}=(u^{i}v^{i})_{x}\, , \qquad i=1,\dots ,N\,,
\end{equation}
where the the `densities' $u^{i}=\eta_i f^i(x,t)$ and the velocities
$v^{i}=-s^i(x,t)$ are related algebraically:
\begin{equation}  \label{02}
v^{i}=\xi _{i}+\sum_{m\neq i}\epsilon _{im}u^{m}(v^{m}-v^{i}) \, , \quad
\epsilon_{ik}=\epsilon_{ki}\, .
\end{equation}
Here
\begin{equation*}
\xi _{i}=-S(\eta _{i})\, ,\qquad \epsilon _{ik}=\frac{1}{\eta _{i}\eta _{k}}%
G(\eta _{i},\eta _{k})\, , \qquad i \ne k \, .
\end{equation*}

Despite the deceptively simple form of system (\ref{01}), (\ref{02}), an
attempt of the analysis of its integrability properties by employing
standard methods of the theory of hydrodynamic type systems (verification of
 the Haantjes tensor vanishing, computation of the Riemann invariants
in terms of the densities of conservation laws, establishing the
semi-Hamiltonian property etc. --- see, e.g. \cite{pss96}) reveals serious
technical problems already for a modest $N=4$. The reason for such unexpected
difficulties in the apparently straightforward procedure lies in the fact
that the existing theory heavily relies on the knowledge of the \textit{%
explicit} dependence of the coefficient matrix of the hydrodynamic type
system on field variables while the dependence $v^i(\mathbf{u})$ in (\ref{01}%
) is given \textit{implicitly} by algebraic system (\ref{02}). It turns out
that the resolution of this system for $v^i$ using standard computer algebra
packages becomes notoriously resource consuming with the growth of $N$ and does
not hold any promise of getting structurally transparent results for the
Riemann invariants and characteristic velocities. This makes the standard
direct route completely prospectless from the viewpoint of proving
integrability of (\ref{01}), (\ref{02}) and obtaining explicit analytic
results \textit{for an arbitrary $N$}. To deal with the specific structure
of system (\ref{01}), (\ref{02}) we develop in this paper a new approach,
which has enabled us to perform the complete analysis of its integrability for
an arbitrary $N$ and, in particular, to derive compact and elegant
representations for the Riemann invariants and characteristic velocities.

\medskip The main results of the paper can be summarized as follows:

\begin{itemize}
\item {We prove that reductions (\ref{01}), (\ref{02}) represent \textit{%
linearly degenerate integrable} systems of hydrodynamic type \textit{for
arbitrary $N$}. This is done by proving the existence of a certain
representation of the densities $u^i$ and velocities $v^i$ in terms of the
so-called St\"ackel matrix which depends on $N$ functions $r^i(x,t)$, which
are the \textit{Riemann invariants} of equations (\ref{01}), (\ref{02}).
 We
also prove that system (\ref{01}), (\ref{02}) belongs to the Egorov class
(see Def. 7.1 in Section 7).} Moreover, as a by-product of our analysis, we
conclude that the system under study is \textit{the only} (up to unessential
transformations) system of hydrodynamic type which is simultaneously Egorov
and linearly degenerate. The characteristic velocities, conservation law
densities and symmetries (commuting flows) for such systems are fixed by $%
N(N-1)/2$ symmetric constants $\epsilon _{ik}$, $i\ne k$, and $N$ constants $%
\xi_i$ (i.e. by $N(N+1)/2$ constants in total).

\item {We derive an explicit Riemann invariant representation of system (\ref%
{01}), (\ref{02}),
\begin{equation}  \label{rim0}
r_{t}^{i}=v^{i}(\mathbf{r})r_{x}^{i},\qquad i=1,\dots ,N\,,
\end{equation}%
where the Riemann invariants $r^{i}$ are expressed in terms of the component
densities $u^{1},u^{2},\dots ,u^{N}$ as
\begin{equation}  \label{ri0}
r^{i}=-\frac{1}{u^{i}}\left( 1+\underset{m\neq i}{\sum }\epsilon
_{im}u^{m}\right) \,,\qquad i=1,\dots ,N
\end{equation}%
and for the characteristic velocities $v^{i}(\mathbf{r})$ we obtain
\begin{equation}
v^{i}=\frac{1}{u^{i}}\sum\limits_{m=1}^{N}\xi _{m}\beta _{im}, \quad %
\hbox{where} \quad u^{i}=\sum\limits_{m=1}^{N}\beta _{im}.  \label{v0}
\end{equation}%
Here the matrix $\boldsymbol{\beta }=-\boldsymbol{\epsilon }^{-1}$} where
the off-diagonal elements of the symmetric matrix $\boldsymbol{\epsilon }$
are fixed by system (\ref{02}) while the diagonal elements are defined as $%
\epsilon _{ii}=r^{i}$.  Remarkably, the off-diagonal symmetric elements of the matrix $%
\boldsymbol{\beta }$ are nothing than the \textit{rotation coefficients} of
the curvilinear conjugate coordinate net associated with system (\ref{rim0}).  We also note that the second formula in (\ref{v0}) is in fact the inversion of the explicit representation (\ref{ri0}). Importantly, the characteristic velocities in (\ref{rim0})
coincide with the transport velocities in the conservation laws (\ref{01}) --- this is the consequence of linear degeneracy
of system (\ref{01}), (\ref{02}).

\item We construct the full set of commuting flows to (\ref{01}), (\ref{02}),
of which $N-2$ are linearly degenerate. This has allowed us, in particular,
to obtain the family of quasi-periodic solutions
\begin{equation}
x+\xi _{i}t=\overset{r^{i}}{\int }\frac{\xi d\xi }{\sqrt{R_{K}(\xi )}}+%
\underset{m\neq i}{\sum }\epsilon _{im}\overset{r^{m}}{\int }\frac{d\xi }{%
\sqrt{R_{K}(\xi )}},\text{ \ }i=1,2,...,N\,,  \label{quasi0}
\end{equation}%
where
\begin{equation*}
R_{K}(\xi )=\overset{K}{\underset{n=1}{\prod }}(\xi -E_{n})\,,
\end{equation*}%
and $E_{1}<E_{2}<\dots <E_{K}$ are real constants ($K=2N+1$ if $N$ is odd
and $K=2N+2$ if $N$ is even)

\item We show that for the special case $N=3$ there exists a family of
similarity solutions to (\ref{rim0}), (\ref{v0}) having the form $\tilde
r^i=t^{-\alpha}l^i(x/t)$, $i=1,2,3$, $\alpha \ne 0$, where each $\tilde r^i$
is a certain rational function of the corresponding Riemann invariant $r^i$ (%
\ref{ri0}) (and hence, is also a Riemann invariant). These solutions are
found in an implicit (hodograph) form. For $N>3$ such solutions generally do
not exist.
\end{itemize}

Integrability, for arbitrary $N$, of the class of the hydrodynamic
reductions studied in this paper is a strong evidence in favour of
integrability of the full nonlocal kinetic equation (\ref{kin1}), at least
for certain choices of the functions $S(\eta) $ and $G(eta, \mu)$ in
the integral closure equation. Of course, such an outcome does not look
surprising for the particular choice (\ref{kdv}) of $S(\eta) $ and $%
G(\eta, \mu)$ corresponding to the thermodynamic limit of the integrable
KdV-Whitham equations but our analysis suggests that the general
integro-differential kinetic equation (\ref{kin1}) is a representative of a
whole new unexplored class of integrable equations with potentially
important physical applications.

The structure of the paper is as follows. In Section 2 we outline the
derivation of the kinetic equations for the gas of the KdV solitons
following the thermodynamic limit procedure of \cite{el03} and extending it
to the entire KdV-Whitham hierarchy. We then introduce the generalised  kinetic
equation (\ref{kin1}), and in Section 3 consider its $N$-component
`cold-gas' hydrodynamic reductions (\ref{01}), (\ref{02}) having the form of
hydrodynamic conservation laws. We then formulate our main Theorem 3.1
stating that the hydrodynamic reductions under study are linearly degenerate
and integrable (in Tsarev's generalised hodograph sense) hydrodynamic type
systems \textit{for any $N$}. Section 4 is devoted to the account of the
main results of the theory of linearly degenerate hydrodynamic type systems.
In Section 5 we prove the statement of the main Theorem 3.1 for the case $%
N=3 $ by explicitly constructing the corresponding St\"ackel matrix and
presenting expressions for the Riemann invariants and characteristic
velocities in terms of the conserved component densities. We also construct
two distinguished families of exact solutions (self-similar and
quasi-periodic) to the $3$-component reduction. In Section 6, the existence
of the Riemann invariant parametrization of the cold-gas hydrodynamic
reduction, via a single St\"ackel matrix, is proved for arbitrary $N$, which
enables us to complete the proof of the main Theorem 3.1 for a general case.
In Section 7, we derive explicit expressions (\ref{ri0}) and (\ref{v0}) for
the Riemann invariants and characteristic velocities in terms of the
component densities. And at last, in Section 8 we derive hydrodynamic
symmetries (commuting flows) of the $N$-component hydrodynamic reductions
under study and then extract the family of linearly degenerate commuting
flows. We conclude in Section 9 with a general outlook and perspectives arising from
our study.

\section{Kinetic equation for a soliton gas as the thermodynamic limit of
the Whitham modulation system}

We start with an outline of the derivation of the kinetic equation for the
gas of the KdV solitons as the thermodynamic limit of the KdV-Whitham system
following \cite{el03}. We then naturally extend this derivation to the entire
Whitham-KdV hierarchy.

Let us consider the Whitham modulation system associated with the $N$-gap
potentials $\phi_N(x,t)$ of the KdV equation (\ref{kdveq}). This system is
most conveniently represented as a single generating equation in the form \cite
{ffm79}:
\begin{equation}
(dp_{N})_{t}=(dq_{N})_{x}\,,  \label{ffm}
\end{equation}%
where $dp_{N}$ and $dq_{N}$ are the quasimomentum and quasienergy
differentials defined on the two-sheeted hyperelliptic Riemann surface of
genus $N$ :
\begin{equation}  \label{gamma}
\Gamma: \quad \mu^{2}(\lambda )=\prod\limits_{j=1}^{2N+1}(\lambda -\lambda
_{j})\,,\qquad \lambda \in {\mathbb{C}},\quad \lambda _{j}\in {\mathbb{R}}\,.
\end{equation}
\begin{equation*}
\lambda_1<\lambda_2< \dots < \lambda_{2N}< \lambda_{2N+1} \, ,
\end{equation*}
with cuts along spectral bands $[\lambda_1, \lambda_2]$, \dots $%
[\lambda_{2j-1}, \lambda_{2j}]$, \dots, $[\lambda_{2N+1}, \infty]$. We
introduce the canonical system of cycles on $\Gamma$ as follows (see
Fig.~1): the $\alpha_j$-cycle surrounds the $j$-th cut clockwise on the
upper sheet, and the $\beta_j$- cycle is canonically conjugated to $\alpha_j$%
's such that the closed contour $\beta_j$ starts at $\lambda_{2j}$ , goes to
$+\infty$ on the upper sheet and returns to $\lambda_{2j}$ on the lower
sheet.
\begin{figure}[h]
\centerline{\includegraphics[scale=0.5]{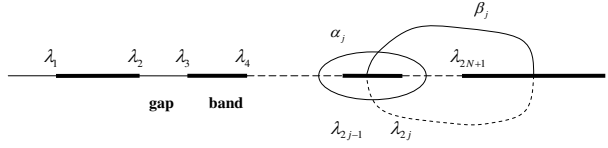}}
\caption{The canonical system of cycles on the hyperelliptic Riemann surface
of genus $N$.}
\end{figure}

The meromorphic differentials $dp_{N}$ and $dq_{N}$ are uniquely defined by
their asymptotic behaviour near $\lambda =-\infty $:
\begin{equation}
-\lambda \gg 1\ :\qquad dp_{N}\sim -\frac{d\lambda }{(-\lambda )^{1/2}}%
\,,\qquad dq_{N}\sim (-\lambda )^{1/2}d\lambda \,  \label{qq}
\end{equation}%
and the normalization
\begin{equation}
\oint\limits_{\beta _{i}}dp_{N}=0\,,\qquad \oint\limits_{\beta
_{i}}dq_{N}=0\,,\qquad i=1,\dots ,N\,;\qquad c_{N}=-\frac{1}{2}%
\sum_{j=1}^{2N+1}\lambda _{j}\,.  \label{norm0}
\end{equation}%
The integrals of $dp_{N}$ and $dq_{N}$ over the $\alpha $ - cycles give the
components of the wave number and the frequency vectors respectively
\begin{equation}
\oint\limits_{\alpha _{j}}dp_{N}(\lambda )=k_{j}(\lambda _{1},\dots ,\lambda
_{2N+1})\,,\qquad \oint\limits_{\alpha _{j}}dq_{N}(\lambda )=\omega
_{j}(\lambda _{1},\dots ,\lambda _{2N+1})\,,\qquad j=1,\dots ,N\,.\
\label{k}
\end{equation}

Let $\lambda_1=-1$, $\lambda_{2N+1}=0$. Following Venakides \cite{ven89} we
introduce a lattice of points
\begin{equation}  \label{lat}
1\approx \eta _1>\eta _2>\ldots >\eta _N\approx 0\, ,
\end{equation}
where
\begin{equation}
-\eta_j^2=\frac 12\left( \lambda_{2j-1}+\lambda_{2j}\right)\,
\end{equation}
are the centres of bands.

We now assume that the spectral bands are distributed such that one can
introduce two positive continuous functions on $[0,1]$:

1. The normalized density of bands $\varphi(\eta)$:

\begin{equation*}
\varphi(\eta)d\eta \approx \frac{\hbox{number of lattice points in} \ (\eta,
\ \eta + d\eta)}{N}\, .
\end{equation*}
That is,
\begin{equation}  \label{phi}
\varphi(\eta_j)= \frac{1}{N(\eta_j - \eta_{j+1})} +O(\frac{1}{N})\, , \qquad
\int \limits_0^1 \varphi(\eta)d\eta =1\, , \ \ \eta^2 = -\lambda \in [0, 1]
\, .
\end{equation}

2. The normalized logarithmic band width $\gamma(\eta)$:
\begin{equation}  \label{gam}
\gamma(\eta_j)= -\frac{1}{N} \log \delta_j+ O(\frac{1}{N})\,, \qquad
\delta_j=\lambda_{2j}-\lambda_{2j-1}\, .
\end{equation}
The functions $\varphi (\eta)$ and $\gamma(\eta)$ asymptotically define the
\textit{local} structure of the Riemann surface $\Gamma$ (\ref{gamma}) for $%
N \gg 1$. In other words, instead of $2N+1$ discrete parameters $\lambda_j$
we have two continuous functions of $\eta$ on $[0,1]$ which do not depend on
$x,t$ on the scale of the typical change of $\lambda_j$'s in (\ref{ffm}),
say $\Delta x \sim \Delta t \sim l$.

The existence of the continuous distributions $\varphi(\eta)$ and $%
\gamma(\eta)$ implies the following band-gap scaling for $N\gg1$:
\begin{equation}  \label{scal}
|\hbox{gap}_j| \sim \frac{1}{\varphi(\eta_j) N} \, , \qquad |\hbox{band}_j|
\sim \exp{\{-\gamma(\eta_j) N \}}\, , \ \ j=1, \dots, N
\end{equation}
Introduction of the distribution (\ref{scal}) is motivated by the structure of the spectrum of
Hill's operator  in the semi-classical limit \cite{wk87}, \cite{ven87}  although
the scaling (\ref{scal}) alone, of course, does not imply exact periodicity of the
(finite-gap) potential.

The scaling (\ref{scal}) has an important property: it preserves the
finiteness of the integrated density of states as $N \to \infty$. The
integrated density of states is defined in terms of the real part of the
quasimomentum integral (see \cite{jm82}):
\begin{equation}  \label{dens0}
\mathcal{N}_N(\lambda)= \frac{1}{\pi} Re \int \limits^{\lambda}_{-1} {%
dp_N(\lambda^{\prime })}\, , \qquad \lambda \in [-1,0]\, .
\end{equation}
Now, using (\ref{k}) one can readily see that
\begin{equation}
\mathcal{N}_N(\lambda) = \frac{1}{2 \pi}\sum _{j=1}^{M} k_j \qquad \hbox{if}
\quad \lambda \in [\lambda_{2M}, \lambda_{2M+1}]\, , \quad M=1, \dots, N\, ,
\end{equation}
which is a particular (finite-gap) case of the general gap-labeling theorem
for quasi-periodic potentials \cite{jm82}. It is not difficult to show that
the scaling (\ref{scal}) implies that $k_j \sim 1/N$ so the total density of
states
\begin{equation}
\mathcal{N}_N(0)= \frac{1}{2 \pi}\sum _{j=1}^{N} k_j
\end{equation}
remains finite in the limit as $N \to \infty$. For this reason we shall call
the continuum limit as $N \to \infty$, defined on the spectral scaling (\ref%
{scal}), the \textit{thermodynamic limit}.

We shall not be concerned here with the existence and the exact meaning of
the thermodynamic limit for the finite-gap potentials $u_N(x,t)$ (which is a
separate interesting problem closely connected with Venakides' continuum limit of theta-functions
 \cite{ven89}) but shall rather directly consider this limit
for the associated Whitham system (\ref{ffm}). It is however, instructive to
note that it follows from (\ref{scal}) that in the thermodynamic limit the
band/gap ratio vanishes for each oscillating mode (i.e. $k_j \to 0$ $\forall
j =1, 2. \dots, N$), so the thermodynamic limit of the sequence of
finite-gap potentials associated with the spectral scaling (\ref{scal}) is
essentially an infinite-soliton limit. It was proposed in \cite{ekmv99} that
this limiting potential should be described in terms of ergodic random
processes and can be viewed as a \textit{homogeneous soliton gas} (or
homogeneous soliton turbulence -- depending on which of the two
``identities" of a soliton is emphasized: the particle or the wave one).
Then it is natural to suppose that the same thermodynamic limit for the
associated Whitham system should describe macroscopic evolution of the
spatially \textit{inhomogeneous} soliton gas. Indeed, as we shall see, the
thermodynamic limit of the Whitham equations turns out to be consistent (in
the small-density limit) with the kinetic equation for solitons derived by
Zakharov \cite{zakh71} using the inverse scattering problem formalism.


We first note that $\mathcal{N}_N(\lambda)$ defined by (\ref{dens0}) is a
monotone increasing positive function so $d\mathcal{N}_N(\lambda)$ is a
measure supported on the spectrum of the finite-gap potential $u_N(x)$ \cite%
{jm82}. Next we introduce a `temporal' analogue of the density of states (%
\ref{dens0}) by the formula
\begin{equation}  \label{dens1}
\mathcal{V}_N(\lambda)= \frac{1}{\pi} Re \int \limits^{\lambda}_{-1} {%
dq_N(\lambda^{\prime })}\, , \qquad \lambda \in [-1,0]\, .
\end{equation}
Then integration of the generating modulation equation (\ref{ffm}) on the
real axis of $\lambda$ from $-1$ to $-\eta^2 \in [-1, 0]$ yields
\begin{equation}  \label{modreal}
\partial_t d \mathcal{N}_N(-\eta^2) = \partial_x d\mathcal{V}_N(-\eta^2) \,
, \qquad \eta \in [0,1].
\end{equation}
Thus the finite-gap Whitham-KdV system can be regarded as the system
governing the evolution of the spectral measure.

Now we consider the thermodynamic limits of $d \mathcal{N}_N$ and $d
\mathcal{V}_N$ which we denote as
\begin{equation}  \label{10}
d \mathcal{N}_N \to \pi f(\eta) d \eta\, , \qquad d\mathcal{V}_N \to -\pi
f(\eta)s(\eta)d\eta\, \qquad \hbox{as} \quad N \to \infty ,
\end{equation}
where the limit is taken on the thermodynamic spectral scaling (\ref{scal}).
Since $\pi f(\eta)d\eta$ is the limiting spectral measure, the function $%
f(\eta)$ has the natural meaning of the distribution function of the
solitons over the spectrum (the meaning of the function $s(\eta)$ will
become clear soon). The functions $f(\eta)$ and $s(\eta)$ were shown in \cite%
{ekmv99}, \cite{el03} to be expressed in terms of the ratio $%
\sigma(\eta)=\varphi(\eta)/\gamma(\eta)$ of the lattice distribution
functions (\ref{phi}), (\ref{gam}) by certain integral equations, which are
then combined into a single equation directly connecting $f(\eta)$ and $%
s(\eta)$ \cite{el03}:
\begin{equation}  \label{s1}
s(\eta)=4\eta^2+\frac{1}{\eta}\int \limits^1_0 \log \left|\frac{\eta + \mu}{%
\eta-\mu}\right|f(\mu)[s(\eta)-s(\mu)]d\mu \, .
\end{equation}
We stress that in the continuum (thermodynamic) limit given by equations (%
\ref{10}), (\ref{s1}) the explicit dependence of the density of states on
the discrete spectral branch points $\lambda_j$ disappears. The only `reminder' of
the hyperelliptic Riemann surface $\Gamma$ (\ref{gamma}) is the kernel $%
\ln|\eta+\mu|/|\eta - \mu|$ which arises as the continuum limit of the
off-diagonal elements of the period matrix $\mathbf{B}$ of the Riemann
theta-function $\Theta_N(x,t|\mathbf{B})$ defining, via the Its-Matveev
formula, the finite-gap potential (see \cite{ven89} and \cite{el03}).

Thus, integral equation (\ref{s1}) can be viewed as a \textit{local} (in the
$x,t$ - plane) relationship between the functions $f(\eta)$ and $s(\eta)$
characterizing the soliton gas. Let $l \gg 1$ be the characteristic length
at which the change of functions $f(\eta)$, $s(\eta)$ is small (of order $%
1/l \ll 1$). Next, in the spirit of the modulation theory (see \cite{wh74}, \cite%
{ffm79}) we assume that on a larger spatiotemporal scale, $\Delta x \gg l$, $%
\Delta t \gg l $, we have $f(\eta) \equiv f(\eta, x,t)$, $s(\eta) \equiv
s(\eta, x,t)$ and postulate, using (\ref{10}), that
\begin{equation}
\partial _t d \mathcal{N}_N \to \pi \partial_t f(\eta, x,t) d \eta\, ,
\qquad \partial _x d \mathcal{V}_N \to -\pi \partial_x [f(\eta, x,t) s(\eta,
x,t)] d \eta\, .
\end{equation}
Then modulation equation (\ref{modreal}) assumes the form of a conservation
equation for $f$,
\begin{equation}  \label{kin11}
f_t+(sf)_x=0\, ,
\end{equation}
which is clearly the expression of the isospectrality of the KdV evolution.
Since $\rho(x,t)=\int^1_0 fd\eta$ is the density of solitons the quantity $%
s(\eta, x,t)$ can naturally be interpreted as the velocity of the soliton
gas (or, more precisely, the velocity of a `trial' soliton with the
spectral parameter $\lambda = -\eta^2$ -- see \cite{gurzyb00}). One can see
from (\ref{s1}) that this velocity differs from the velocity $4 \eta^2$ of
the free soliton with the same spectral parameter. This difference is
obviously due to the collisions of the `trial' $\eta$-soliton with other `$%
\mu$' - solitons in the soliton gas. Indeed, for small densities $\rho=\int
f d\eta \ll 1$ one can consider the second term in (\ref{s1}) as a small
correction to the free-soliton velocity and obtain that to the first order in $
\rho$
\begin{equation}  \label{s1z}
s(\eta) \approx 4\eta^2+\frac{1}{\eta}\int \limits^{1}_0 \ln \left|\frac{%
\eta + \mu}{\eta-\mu}\right|f(\mu)[4\eta^2-4\mu^2]d\mu \, ,
\end{equation}
which is Zakharov's expression for the average velocity of a trial soliton
in a rarefied soliton gas, obtained in \cite{zakh71} by taking into account the change in the
soliton position due to phase shifts in its pairwise collisions with other
solitons.  We would like to emphasize crucial difference between the mathematical structure of Zakharov's
 asymptotic formula (\ref{s1z}), which represents an {\it explicit expression} for the trial soliton velocity $s(\eta)$ in terms of
 the spectral distribution function $f(\eta)$, and that of formula (\ref{s1}) which is a non-perturbative {\it integral equation}  for $s(\eta)$.

Equations (\ref{kin11}) and (\ref{s1}) thus provide a self-consistent
kinetic description of the KdV soliton gas of finite density. We note that
the upper limit in the integrals in (\ref{s1}), (\ref{s1z}) can be replaced
by $+\infty$ to make the kinetic equation independent on the original
spectral lattice normalization (\ref{lat}).

The outlined procedure of the thermodynamic limit can be readily extended to
the entire Whitham-KdV hierarchy,
\begin{equation}
(dp_{N})_{t_{n}}=(dq_{N}^{(n)})_{x}\,,\quad n\in \mathbb{N}\,,  \label{ffmc}
\end{equation}%
where $n$ is the number of the \textquotedblleft higher\textquotedblright\
Whitham-KdV equation in the hierarchy (the original modulation equation (\ref%
{ffm}) corresponding to the KdV equation itself has the number $n=1$) and $%
t_{n}$ is the corresponding \textquotedblleft higher" time, so that $%
(dp_{N})_{t_{n}t_{m}}=(dp_{N})_{t_{m}t_{n}}$ for all $n\neq m$. The
meromorphic differential $dq_{N}^{(n)}$ is uniquely defined by its
asymptotic behaviour near $\lambda =-\infty $,
\begin{equation}
dq^{(n)}\sim (-\lambda )^{n-1/2}d\lambda \,,  \label{qqn}
\end{equation}%
and the normalization
\begin{equation}
\oint\limits_{\beta _{j}}dq^{(n)}(\lambda )=0\,,\qquad j=1,\dots ,N\,
\label{norm1}
\end{equation}%
analogous to (\ref{norm0}).

Now, applying  the  above procedure of the
thermodynamic limiting transition to equation (\ref{ffmc}) we obtain the same transport equation (\ref%
{kin11}) for the distribution function $f(\eta, x,t)$
\begin{equation}  \label{kin2}
f_{t_n}+(s_nf)_x=0\, ,
\end{equation}
while the integral closure equation for $s_n$ assumes the form
\begin{equation}  \label{s2}
s_n(\eta)=C_n\eta^{2n}+\frac{1}{\eta}\int \limits^1_0 \log \left|\frac{\eta
+ \mu}{\eta-\mu}\right|f(\mu)[s_n(\eta)-s_n(\mu)]d\mu \, ,
\end{equation}
where $C_n$ are certain constants whose specific values won't be required below.
Moreover, since the characteristic speeds of the commuting
KdV-Whitham flows, and, therefore, the corresponding transport velocities $%
s_n$ in the thermodynamic limit equation (\ref{s2}), are defined up to a
constant factor, hereafter one can assume $C_n$ to be arbitrary constants.

We note that equation (\ref{s2}) differs from (\ref{s1}) only in the first
term corresponding to the free-soliton velocity. Also note that the
`phase-shift' logarithmic kernel in the integral equation (\ref{s2}) is the
same for all $n$, which is not surprising as the  entire finite-gap
Whitham-KdV hierarchy (\ref{ffmc}) is associated with the same Riemann
surface, i.e. with the same period matrix $\mathbf{B}$ responsible for the
form of the integral kernel in the limit.

Now it is only natural to consider a generalization of the derived kinetic
equations (\ref{kin2}), (\ref{s2}) by introducing in (\ref{s2}) an arbitrary
function $S(\eta)$ instead of the free-soliton velocity term and an arbitrary symmetric
function $G(\eta, \mu)$ instead of the logarithmic `phase-shift kernel' in the integral
term. Also, as was already mentioned, we replace the upper limit of integration in the
closure equation (\ref{s1}) by $+ \infty$. As a result, we arrive at the
generalised kinetic equation (\ref{kin1}), which will be our main concern in
the remainder of the paper.

\section{`Cold-gas' hydrodynamic reductions}

We introduce
an $N$-component `cold-gas' \textit{ansatz} for the distribution function $f(\eta,x,t)$:
\begin{equation}
f=\sum\limits_{i=1}^{N}f^{i}(x,t)\delta (\eta -\eta _{i})\,,  \label{delta}
\end{equation}%
where $\eta _{N}>\eta _{N-1}>\dots >\eta _{1}>0$ are arbitrary numbers and $f^i(x,t)$, $n=1, \dots, N$ are unknown functions.

Before we proceed with the analysis of  mathematical consequences of this `cold-gas' ansatz
it is instructive to say a couple of words about its physical meaning (see \cite{elkam05}). To be definite, we
shall refer to the KdV case here. The distribution (\ref{delta}) represents an idealized description of the distribution function in a soliton gas with the solitons having their
spectral parameters $\eta$ distributed in narrow vicinities of $N$ discrete values $\eta_i$.
As a matter of fact, owing to non-degeneracy of discrete spectrum of the linear Schr\"odinger operator, all individual spectral  parameters within the $i$-th  component of the soliton gas must be different.
The soliton positions in such a `quasi-monochromatic' component of the soliton gas  are statistically independent
which results in the Poisson distribution with the mean density $f_i$ for the number of solitons in
a unit space interval (the Poisson distribution
naturally arises in the thermodynamic limit of finite-gap potentials  \cite{ekmv99}).
It is also clear that one can neglect the effect of the interactions between the
solitons belonging to the same gas component compared with the cross-component
interactions (the typical time of the interactions between solitons with close values of the spectral parameter is much larger than when these parameters are mutually spaced within the spectral interval --- see, e.g., \cite{novTS}). This will be shown below to have important mathematical consequences.

Substitution (\ref{delta}) reduces (\ref{kin1}) to a system of hydrodynamic
conservation laws,
\begin{equation}
u_{t}^{i}=(u^{i}v^{i})_{x},\qquad i=1,\dots ,N\,,  \label{cont}
\end{equation}%
where the component `densities' $u^i$ and the velocities $v^i$
defined as
\begin{equation}\label{dv}
u^i(x,t)=\eta_i f^i(x,t)\, , \qquad v^{i}(x,t)=-s(\eta _{i},x,t)\, ,
\end{equation}
are related algebraically
\begin{equation}
v^{i}=\xi _{i}+\sum_{m\neq i}\epsilon _{im}u^{m}(v^{m}-v^{i})\text{, \ \ \ \
}\epsilon _{ik}=\epsilon _{ki}.  \label{alg}
\end{equation}%
Here
\begin{equation}
\xi _{i}=-S(\eta _{i})\,,\qquad \epsilon _{ik}=\frac{1}{\eta _{i}\eta _{k}}%
G(\eta _{i},\eta _{k})\,\quad i\neq k\,.  \label{rel1}
\end{equation}%
Note that the quantities $\epsilon_{ii}$ are not defined.

In a two-component case, the above algebraic system (\ref{alg}) can be
easily resolved for $u^{1,2}$ in terms of $v^{1,2}$:
\begin{equation}  \label{alg2}
u^{1}=\frac{1}{\epsilon _{12}}\frac{v^{2}-\xi _{2}}{v^{1}-v^{2}}\text{, \ \
\ \ \ \ }u^{2}=\frac{1}{\epsilon _{12}}\frac{v^{1}-\xi _{1}}{v^{2}-v^{1}}.
\end{equation}
Substituting (\ref{alg2}) into (\ref{cont}) we arrive at the

\textbf{Lemma 3.1} (El \& Kamchatnov 2005 \cite{elkam05}): \textit{%
Hydrodynamic type system (\ref{cont} ), (\ref{alg}) for $N=2$ reduces to a
diagonal form in the field variables $v^{1}$ and $v^{2}$}:
\begin{equation}
v_{t}^{1}=v^{2}v_{x}^{1}\text{, \ \ \ \ \ }v_{t}^{2}=v^{1}v_{x}^{2}.
\label{chap}
\end{equation}%
Remarkably, the hydrodynamic type system (\ref{chap}) is \textit{linearly
degenerate} because its characteristic velocities do not depend on the
corresponding Riemann invariants.  Physically this linear degeneracy reflects the already mentioned domination of the `cross-component'
soliton interactions over the interactions within a given component consisting of solitons with close amplitudes.

It is worth noting that system (\ref{chap}) is equivalent to the 1D
Born-Infeld equation (Born \& Infeld 1934) arising in nonlinear
electromagnetic field theory (see \cite{wh74}, \cite{arnut89})
\begin{equation*}
(1+\varphi _{x}^{2})\varphi _{yy}-2\varphi _{x}\varphi _{y}\varphi
_{xy}+(1-\varphi _{y}^{2})\varphi _{xx}=0.
\end{equation*}

As any two-component hydrodynamic type system, (\ref{chap}) is integrable
(linearizable) via the classical hodograph transform. However, for any $%
N\geq 3$ integrability of the original system (\ref{cont}), (\ref{alg}) is
no longer obvious. As a matter of fact, most $N$-component hydrodynamic type
systems are \textit{not integrable} for $N\geq 3$. Also, it is even not
clear whether $N$-component system (\ref{cont}), (\ref{alg}) is linearly
degenerate. It might seem that this system is simple enough for one to be
able to verify these properties by a direct computation, using general
definitions of linear degeneracy and integrability for hydrodynamic type
systems \cite{pav87}, \cite{ts85, ts91} (also see the next Section). To our
surprise, even the simplest non-trivial case $N=3$ turned out to be
complicated enough to require computer algebra to get the confirmation of
our hypothesis.

The identification of the system (\ref{cont}), (\ref{alg}) for $N=3$ as an
integrable linearly degenerate hydrodynamic system can be considered as a
strong indication that both properties (linear degeneracy and integrability)
could hold true for this system for arbitrary $N$. Thus we formulate our main

\medskip

\textbf{Theorem 3.1}\ \textit{$N$-component reductions (\ref{cont}), (\ref%
{alg}) of the generalised kinetic equation (\ref{kin1}) are linearly
degenerate integrable hydrodynamic type systems for any $N$.}

\medskip To prove this theorem, we take advantage of the well-developed
theory of integrable (semi-Hamiltonian) linearly degenerate hydrodynamic type systems \cite%
{pav87}, \cite{fer91}. For convenience, in the next section we present a brief review of the main results of this theory which will be extensively used in Sections 5 -- 8 of the paper.

\section{Linearly degenerate integrable hydrodynamic type systems: account
of properties}


A hydrodynamic type system%
\begin{equation}
U_{t}^{i}=v_{j}^{i}(\mathbf{U})U_{x}^{j},\text{ \ \ \ \ \ }i,j=1,2,...,N
\label{gen}
\end{equation}%
is called semi-Hamiltonian (see \cite{ts85, ts91}) if it

\textbf{(i)} has $N$ mutually distinct eigenvalues $\lambda = \lambda ^{i}(\mathbf{U})$ defined by the equation
\begin{equation}
\det \left\vert \lambda \delta _{j}^{i}-v_{j}^{i}(\mathbf{U})\right\vert =0;
\label{det}
\end{equation}

\textbf{(ii)} admits invertible point transformations $U^{k}(\mathbf{r})$,
such that this hydrodynamic type system can be written in diagonal form%
\begin{equation}
r_{t}^{i}=V^{i}(\mathbf{r})r_{x}^{i},\qquad i=1,\dots ,N.  \label{rim}
\end{equation}%
The variables $r^{k}(\mathbf{U})$ are called Riemann invariants and $V^{k}(%
\mathbf{r})=\lambda ^{k}(\mathbf{U}(\mathbf{r}))$ -- characteristic
velocities. Each Riemann invariant $r^{i}$ is determined up to an arbitrary
function of a single variable $R_{i}(r^{i})$.

\textbf{(iii)} satisfies the identity%
\begin{equation}
\partial _{j}\frac{\partial _{k}V^{i}}{V^{k}-V^{i}}=\partial _{k}\frac{%
\partial _{j}V^{i}}{V^{j}-V^{i}}\text{, \ \ \ }i\neq j\neq k  \label{iden}
\end{equation}%
for each three distinct characteristic velocities ($\partial _{k}\equiv
\partial /\partial r^{k}$).

A semi-Hamiltonian hydrodynamic type system possesses infinitely many
conservation laws parameterised by $N$ arbitrary functions of a single
variable. Its general local solution for $\partial _{x}r^{i}\neq 0$, $%
i=1,\dots ,N$ is given by the generalised hodograph formula \cite{ts85, ts91}
\begin{equation}
x+V^{i}(\mathbf{r})t=W^{i}(\mathbf{r)\,,}  \label{hod}
\end{equation}
where functions $W^{i}(\mathbf{r})$ are found from the linear system of
PDEs:
\begin{equation}
\frac{\partial _{i}W^{j}}{W^{i}-W^{j}} = \frac{\partial _{i}V^{j}}{%
V^{i}-V^{j}} \,,\quad i,j=1,\dots ,N,\quad i\neq j.  \label{ts}
\end{equation}
Thus, the semi-Hamiltonian property (\ref{iden}) implies integrability of
diagonal hydrodynamic type system in the above generalised hodograph sense.

It is known \cite{ts91} that solutions $W^j$ of (\ref{ts}) specify commuting hydrodynamic flows to (\ref{rim}%
):
\begin{equation}
r_{\tau }^{j}=W^{j}(\mathbf{r})r_{x}^{j}\,,\quad j=1,\dots ,N \, ,
\label{comm}
\end{equation}%
where $\tau $ is a new time (group parameter).  Indeed, one can readily show that equations (\ref{rim}), (\ref{comm}), (\ref{ts}) imply $(r_{\tau
}^{j})_{t}=(r_{t}^{j})_{\tau }$.

A sub-class of linearly degenerate hydrodynamic type systems is distinguished
by the property
\begin{equation}
\partial _{i}V^{i}=0  \label{lin}
\end{equation}%
for each index $i$. It means that each characteristic velocity does not
depend on the corresponding Riemann invariant $r^{i}$.

\medskip \textbf{Theorem 4.1} (Pavlov 1987 \cite{pav87}): \textit{\ If
semi-Hamiltonian hydrodynamic type system (\ref{rim}) possesses conservation
laws (\ref{cont}) with $u^i=U^i({\mathbf{r}})$ and $v^{i}(\mathbf{U(r)}%
)=V^{i}(\mathbf{r})$ then this system is linearly degenerate. These
conservation laws (\ref{cont}) are parameterised by $N$ arbitrary functions
of a single variable.}

\medskip

\textbf{Proof}: The semi-Hamiltonian property (i.e. \textit{integrability})
is given by the condition (\ref{iden}). We introduce, following Tsarev \cite%
{ts91}, the so-called Lame coefficients $\bar{H}_{i}$ by
\begin{equation}
\partial _{k}\ln \bar{H}_{i}=\frac{\partial _{k}V^{i}}{V^{k}-V^{i}}\text{, \
\ \ }i\neq k.  \label{lam}
\end{equation}%
Suppose that some semi-Hamiltonian hydrodynamic type system (\ref{rim}) can
be written in the conservative form (\ref{cont}) with $v^{i}(\mathbf{U(r)}%
)=V^{i}(\mathbf{r})$. In such a case%
\begin{equation*}
\partial _{k}U^{i}\cdot r_{t}^{k}=\partial _{k}(U^{i}V^{i})\cdot r_{x}^{k}.
\end{equation*}%
Since $\mathbf{r}(x,t)$ is an arbitrary solution of (\ref{rim}) we obtain $N$
equations%
\begin{equation}
V^{k}\cdot \partial _{k}U^{i}=\partial _{k}(U^{i}V^{i}).  \label{18}
\end{equation}%
If $k\neq i$, then%
\begin{equation}
\partial _{k}\ln U^{i}=\frac{\partial _{k}V^{i}}{V^{k}-V^{i}},  \label{den}
\end{equation}%
i.e. each of the conservation law densities $U^{i}$ is determined up to an
arbitrary function of a single variable $P_{i}(r^{i})$ (cf. (\ref{lam}) and (%
\ref{den})),
\begin{equation}
U^{i}=\bar{H}_{i}\cdot P_{i}(r^{i}).  \label{uH}
\end{equation}%
If $k=i$, then it follows from (\ref{18}) that $\partial _{i}V^{i}=0$ i.e.
the system is linearly degenerate. The theorem is proved.

\medskip

\textbf{Remark 1}: A subset $\{u^{k}\}$ of the conservation law densities $%
\{U^k\}$ satisfying a given system of conservation laws (e.g. (\ref{cont}), (%
\ref{alg})) is selected by fixing the functions $P_k$ (e.g. $P_{k}(r^{k})
\equiv 1$ --- see (\ref{norm}) in Section 7).

\medskip

 While converse of Theorem 4.1 is also true, one should note that not {\it every} conservation law of a semi-Hamiltonian
linearly degenerate system satisfies the key property $v^i=V^i$. Indeed, let us consider the two-component system of conservation
laws,%
\begin{equation}
U_{t}^{1}=(U^{1}v^{1}(U^{1},U^{2}))_{x}\text{, \ \ \ \ }%
U_{t}^{2}=(U^{2}v^{2}(U^{1},U^{2}))_{x}.  \label{cons12}
\end{equation}%
Suppose this hydrodynamic type system is linearly degenerate, then it can be
written in Riemann invariants $r^{1}(U^{1},U^{2})$, $r^{2}(U^{1},U^{2})$ as
follows:
\begin{equation*}
r_{t}^{1}=V^{1}(r^{1},r^{2})r_{x}^{1}\text{, \ \ \ }%
r_{t}^{2}=V^{2}(r^{1},r^{2})r_{x}^{2},
\end{equation*}%
where $V^{1,2}(\mathbf{r})=v^{1,2}(\mathbf{U}(\mathbf{r}))$. Let us
introduce new conservation law densities $\tilde{U}^{1}=U^{1}+U^{2}$ and $%
\tilde{U}^{2}=U^{1}-U^{2}$. Then the system of conservation laws (\ref%
{cons12}) assumes an equivalent form%
\begin{equation*}
\tilde{U}_{t}^{1}=(\tilde{U}^{1}\tilde{v}^{1}(\tilde{U}^{1},\tilde{U}%
^{2}))_{x}\text{, \ \ \ \ }\tilde{U}_{t}^{2}=(\tilde{U}^{2}\tilde{v}^{2}(%
\tilde{U}^{1},\tilde{U}^{2}))_{x},
\end{equation*}%
where the characteristic velocities%
\begin{equation*}
\tilde{v}^{1}=\frac{U^{1}v^{1}+U^{2}v^{2}}{U^{1}+U^{2}}\text{, \ \ \ }\tilde{%
v}^{2}=\frac{U^{1}v^{1}-U^{2}v^{2}}{U^{1}-U^{2}}
\end{equation*}%
no longer coincide with $V^{1}(r^{1},r^{2})$ and $V^{2}(r^{1},r^{2})$.

\medskip

The full theory of linearly degenerate semi-Hamiltonian hydrodynamic type
systems was constructed by Ferapontov in \cite{fer91} using the St\"{a}ckel
matrices
\begin{equation}
\Delta =\left(
\begin{array}{ccccc}
\phi _{1}^{1}(r^{1}) &  & ... &  & \phi _{N}^{1}(r^{N}) \\
&  &  &  &  \\
... &  & ... &  & ... \\
&  &  &  &  \\
\phi _{1}^{N-2}(r^{1}) &  &  &  & \phi _{N}^{N-2}(r^{N}) \\
\phi _{1}^{N-1}(r^{1}) &  &  &  & \phi _{N}^{N-1}(r^{N}) \\
1 &  & ... &  & 1%
\end{array}%
\right)  \label{stak}
\end{equation}%
where $\phi _{k}^{i}(r^{k})$ are $N(N-1)$ arbitrary functions (it is clear
that without loss of generality one can put $\phi _{k}^{N-1}(z)=z$ and the
number of arbitrary function reduces to $N(N-2)$). Then the characteristic
velocities of such linearly degenerate hydrodynamic type systems are given
by the formula
\begin{equation}
V^{i}=\frac{\det \Delta _{i}^{(2)}}{\det \Delta _{i}^{(1)}}\,,  \label{vstak}
\end{equation}%
where $\Delta _{i}^{(k)}$ is the matrix $\Delta $ without $k$th row and $i$%
th column. The family of the conservation law densities $U^i$ corresponding
to the semi-Hamiltonian system (\ref{rim}), (\ref{vstak}) is determined by
(cf. (\ref{uH}))
\begin{equation}
U^{i}=\frac{\det \Delta _{i}^{(1)}}{\det \Delta }(-1)^{i+1}P_{i}(r^{i}),
\label{FerSbs}
\end{equation}%
where $P_{i}(r^{i})$, $i=1,\dots ,N$ are arbitrary functions. \newline

\textbf{Corollary 4.1} \textit{The system of conservation laws (\ref{cont})
is a semi-Hamiltonian linearly degenerate hydrodynamic type system if and
only if the densities $u^{i}$ and velocities $v^{i}(\mathbf{u})$ admit
representations $u^i=U^i({\mathbf{r}})$ and $v^{i}(\mathbf{U(r)})=V^{i}(%
\mathbf{r})$, specified by (\ref{FerSbs}), (\ref{vstak}), via $N$ functions $%
r^{k}(x,t)$ satisfying diagonal system (\ref{rim}).} \newline

\medskip \textbf{Proposition 4.1} (Ferapontov 1991 \cite{fer91}): \textit{%
Semi-Hamiltonian linearly degenerate hydrodynamic type system (\ref{rim}), (%
\ref{vstak}) has $N-2$ nontrivial \underline{linearly degenerate} commuting
flows}
\begin{equation}
r_{t^{k}}^{j}=V_{(k)}^{j}(\mathbf{r})r_{x}^{j},\quad j=1,\dots ,N\,,\quad
k=3,4,...,N,  \label{commk}
\end{equation}%
whose characteristic velocities are determined as (cf. (\ref{vstak}))
\begin{equation}
V_{(k)}^{i}=\frac{\det \Delta _{i}^{(k)}}{\det \Delta _{i}^{(1)}}\,.
\label{deta}
\end{equation}%
Any characteristic velocity vector $\boldsymbol{\mathcal{W}}(\mathbf{r})=(%
\mathcal{W}^{1}(\mathbf{r}),\mathcal{W}^{2}(\mathbf{r}),\dots ,\mathcal{W}%
^{N}(\mathbf{r}))$ specifying linearly degenerate hydrodynamic flow $r_{\tau
}^{j}=\mathcal{W}^{j}(\mathbf{r})r_{x}^{j}\,,\ j=1,\dots ,N$, commuting with
(\ref{rim}), (\ref{vstak}), can be represented as a linear combination of
the \textquotedblleft basis\textquotedblright\ characteristic velocity
vectors $\mathbf{V}_{(k)}$ (\ref{deta}) (including \textquotedblleft
trivial\textquotedblright\ ones $\mathbf{V}_{(2)}\equiv \mathbf{V}$ (see (%
\ref{vstak})) and $\mathbf{V}_{(1)}\equiv \mathbf{1}$) with some constant
coefficients $b_{k}$. Thus, for any component $\mathcal{W}^{i}$ there exists
a decomposition
\begin{equation}
\mathcal{W}^{i}=\sum\limits_{k=1}^{N}b_{k}V_{(k)}^{i}\,.  \label{decomp}
\end{equation}

\medskip

\textbf{Theorem 4.2} (Ferapontov 1991 \cite{fer91}): \textit{General
solution $\mathbf{r}(x,t)$ of the semi-Hamiltonian linearly degenerate
system (\ref{rim}) is parameterised by $N$ arbitrary functions of one
variable $f_{k}(r^{k})$ and is given in an implicit form by the algebraic
system}
\begin{eqnarray}
x &=&\sum_{k=1}^{N}\overset{r^{k}}{\int }\frac{\phi _{k}^{1}(\xi )d\xi }{%
f_{k}(\xi )}\text{, \ \ \ \ }-t=\sum_{k=1}^{N}\overset{r^{k}}{\int }\frac{%
\phi _{k}^{2}(\xi )d\xi }{f_{k}(\xi )}  \notag \\
&&  \label{theo} \\
0 &=&\sum_{k=1}^{N}\overset{r^{k}}{\int }\frac{\phi _{k}^{M}(\xi )d\xi }{%
f_{k}(\xi )}\text{, \ \ \ \ }M=3,4,...,N.  \notag
\end{eqnarray}%
(note the change of sign for $t$ compared to \cite{fer91} due to a slightly
different representation of the diagonal system (\ref{rim}) in this paper).
We also note that formulae (\ref{theo}) represent an equivalent of the
symmetric generalised hodograph solution (\ref{hod}) for semi-Hamiltonian
linearly degenerate hydrodynamic type systems.

It is instructive to introduce, following Darboux \cite{darboux}, the
so-called rotation coefficients%
\begin{equation}
\beta _{ik}=\frac{\partial _{i}\bar{H}_{k}}{\bar{H}_{i}}\text{, \ \ \ }i\neq
k\,,  \label{rot}
\end{equation}%
where the Lam\'{e} coefficients $\bar{H}_{i}$ are defined by (\ref{lam}).
Then expression (\ref{iden}) for the semi-Hamiltonian property assumes the
form of a Darboux system
\begin{equation}
\partial _{i}\beta _{jk}=\beta _{ji}\beta _{ik},\text{\ \ \ }i\neq j\neq k\,.
\label{55}
\end{equation}%
Using (\ref{rot}) linear system (\ref{ts}) can be related to another linear
system
\begin{equation}
\partial _{i}H_{k}=\beta _{ik}H_{i},\text{\ \ \ }i\neq k,  \label{comb}
\end{equation}
via the so-called Combescure transformation (see \cite{darboux})%
\begin{equation}
W^{i}=\frac{H_{i}}{\bar{H}_{i}}.  \label{coma}
\end{equation}%
In other words, one can show (see \cite{ts91}) that the ratio of any two
solutions to (\ref{comb}) satisfies system (\ref{ts}) for the characteristic
velocities of the commuting flows (\ref{comm}). We note that the particular
solution $\tilde{H}_{i}$ of (\ref{comb}) corresponding to the characteristic
velocities $V_{i}$ of the original system (\ref{rim}) is expressed in terms
of the Lam\'{e} coefficient $\bar{H}_{i}$ as
\begin{equation}
\tilde{H}_{i}=V^{i}\bar{H}_{i}.  \label{sec}
\end{equation}

Of course, general solution $H_{i}$ of system (\ref{comb}), as well as
general solution $W^{i}$ of the generalised hodograph equations (\ref{ts}),
is parameterised by $N$ arbitrary functions of a single variable.

\textbf{Theorem 4.3} (Pavlov 1987 \cite{pav87}): \textit{The class of the
semi-Hamiltonian linearly degenerate systems of hydrodynamic type is
selected, in addition to (\ref{55}), by the set of restrictions on the
rotation and Lame coefficients}%
\begin{equation}
\partial _{i}\ln \bar{H}_{i}=\partial _{i}\ln \beta _{ji}  \label{ld}
\end{equation}%
\textit{for any index} $j\neq i$.

\textbf{Proof}: Let us consider the Lam\'{e} coefficients for the linearly
degenerate systems. Using (\ref{lam}), (\ref{lin}) we have
\begin{equation*}
\partial _{j}V^{i}=\partial _{j}\ln \bar{H}_{i}\cdot (V^{j}-V^{i}),\text{ \ }%
i\neq j,\text{ \ \ \ \ \ }\partial _{i}V^{i}=0.
\end{equation*}%
The compatibility condition $\partial _{i}(\partial _{j}V^{i})=\partial
_{j}(\partial _{i}V^{i})$ implies that
\begin{equation}
\partial _{i}\partial _{j}\ln \bar{H}_{i}=\partial _{j}\ln \bar{H}_{i}\cdot
\partial _{i}\ln \bar{H}_{j},\text{\ \ \ }i\neq j.  \label{comp}
\end{equation}%
Now one can see that the l.h.s. of (\ref{comp}) can be written in the form%
\begin{equation}
\partial _{i}\partial _{j}\ln \bar{H}_{i}=\partial _{i}\left( \frac{\bar{H}%
_{j}}{\bar{H}_{i}}\beta _{ji}\right) =\beta _{ij}\beta _{ji}+\frac{\bar{H}%
_{j}}{\bar{H}_{i}}\partial _{i}\beta _{ji}-\frac{\bar{H}_{j}}{\bar{H}_{i}^{2}%
}\beta _{ji}\partial _{i}\bar{H}_{i}.  \label{lhs}
\end{equation}%
On the other hand, the r.h.s. of (\ref{comp}) is nothing but the product $%
\beta _{ij}\beta _{ji}$. Now (\ref{ld}) immediately follows from (\ref{comp}%
) and (\ref{lhs}). The Theorem is proved.

Now, suppose that the rotation coefficients (\ref{rot}) for some linearly
degenerate hydrodynamic type system are given. Then restrictions (\ref{ld})
determine not only the Lam\'{e} coefficients (\ref{lam}) but also all other
solutions of (\ref{comb}) associated, via (\ref{coma}), with the
characteristic velocities (\ref{vstak}), (\ref{deta}) of the complete set of
linearly degenerate commuting flows. Indeed, one can see that equations (\ref%
{rot}), (\ref{ld}) actually represent $N$ systems of \textit{ordinary
differential equations} so that each system contains differentiation with
respect to only one Riemann invariant. Thus, the general solution $\bar{H}%
_{i}$ of system (\ref{rot}), (\ref{ld}) is parameterised by $N$ arbitrary
constants (see Proposition 4.1).

Let us introduce $N$ particular solutions $\bar{H}_{i}^{(k)}$ of system (\ref%
{rot}), (\ref{ld}) such that (see (\ref{vstak}), (\ref{deta}))%
\begin{equation*}
V_{(k)}^{i}=\frac{\bar{H}_{i}^{(k)}}{\bar{H}_{i}},\text{ \ }k=1,2,...,N,
\end{equation*}%
where $\bar{H}_{i}=\bar{H}_{i}^{(1)},\tilde{H}_{i}=\bar{H}_{i}^{(2)}$ (see (%
\ref{deta})). As a matter of fact, $V_{(2)}^{i}\equiv
V^{i},V_{(1)}^{i}\equiv 1$. Then (\ref{ld}) can be written in a slightly
more general form,
\begin{equation*}
\partial _{i}\ln \beta _{ji}=\partial _{i}\ln \bar{H}_{i}^{(k)}\,,
\end{equation*}%
-- for any $k$ and $j\neq i$.

Thus, the full class of linearly degenerate semi-Hamiltonian hydrodynamic
type systems is determined by conditions (\ref{ld}), (\ref{rot}) and (\ref%
{55}). We note that system (\ref{ld}), (\ref{rot}) and (\ref{55}) is an
overdetermined system in involution. Its integration leads to the
aforementioned set of particular solutions of (\ref{comb}) that can be
parameterised via a St\"{a}ckel matrix (see (\ref{stak}), (\ref{vstak}), (%
\ref{FerSbs}) and (\ref{deta})).

\section{$N=3$: explicit formulae}

We now consider the first nontrivial (from the viewpoint of integrability)
case $N=3$ of the hydrodynamic reduction (\ref{cont}), (\ref{alg}). To prove
our main 
Theorem 3.1 for $N=3$ we shall make use of Corollary 4.1.

Let us suppose that hydrodynamic system of conservation laws (\ref{cont}), (%
\ref{alg}) is linearly degenerate and can be written in a diagonal form (\ref%
{rim}), i.e. we suppose that there exists an invertible change of variables $%
r^{j}(\mathbf{u})\,,\ j=1,2,3,$ such that system (\ref{cont}) assumes a
diagonal form%
\begin{equation}
r_{t}^{j}=V^{j}(\mathbf{r})r_{x}^{j}\,,\qquad j=1,2,3,  \label{R3}
\end{equation}%
where $V^{j}(\mathbf{r})=v^{j}(\mathbf{u(r)})$.

We introduce the St\"{a}ckel matrix (\ref{stak}), which for $N=3$ can be
written in the form
\begin{equation}
\Delta = \left(
\begin{array}{ccc}
B_{1}(r^{1}) & B_{2}(r^{2}) & B_{3}(r^{3}) \\
A_{1}(r^{1}) & A_{2}(r^{2}) & A_{3}(r^{3}) \\
1 & 1 & 1%
\end{array}%
\right) ,  \label{st}
\end{equation}%
where $A_{k}(z)$ and $B_{k}(z)$ are arbitrary functions.

Now, by Corollary 4.1, if system (\ref{cont}), (\ref{alg}) is linearly
degenerate and semi-Hamiltonian then its diagonal representation (\ref{R3})
must have characteristic velocities in the form (\ref{vstak}), i.e. for $N=3$
we have
\begin{equation}
V^{1}=\frac{B_{2}(r^{2})-B_{3}(r^{3})}{A_{2}(r^{2})-A_{3}(r^{3})}\text{, \ \
\ }V^{2}=\frac{B_{3}(r^{3})-B_{1}(r^{1})}{A_{3}(r^{3})-A_{1}(r^{1})}\text{,
\ \ \ }V^{3}=\frac{B_{1}(r^{1})-B_{2}(r^{2})}{A_{1}(r^{1})-A_{2}(r^{2})}.
\label{r3}
\end{equation}%
Then, using (\ref{FerSbs}) the corresponding conservation law densities $%
u^{k}$ are found in terms of Riemann invariants as
\begin{equation}
u^{1}=\frac{P_{1}(r^{1})}{\det \Delta}[A_{2}(r^{2})-A_{3}(r^{3})]\text{, \ \
}u^{2}=\frac{P_{2}(r^{2})}{\det \Delta}[A_{3}(r^{3})-A_{1}(r^{1})]\text{, \
\ }u^{3}=\frac{P_{3}(r^{3})}{\det \Delta}[A_{1}(r^{1})-A_{2}(r^{2})],
\label{u3}
\end{equation}%
where $P_{j}(r^{j})$ are arbitrary functions and the determinant of the St%
\"{a}ckel matrix is given by
\begin{equation}
\det \Delta
=A_{1}(r^{1})[B_{2}(r^{2})-B_{3}(r^{3})]+A_{2}(r^{2})[B_{3}(r^{3})-
B_{1}(r^{1})]+A_{3}(r^{3})[B_{1}(r^{1})-B_{2}(r^{2})].  \label{det3}
\end{equation}

Substitution of (\ref{r3})--(\ref{det3}) into (\ref{alg}) yields expressions
for the functions $A_{k}(z)$, $B_{k}(z)$, $P_k(z)$, $k=1,2,3$.

Before we present these expressions, we note that it follows from (\ref{r3}%
), (\ref{det3}) that functions $B_{k}(z)$ are determined up to a constant
shift which is then translated into a certain shift for the functions $%
P_{k}(z)$. It turns out that this shift can be removed by the simplest
change of the Riemann invariants, $(r^{k}+{\hbox{constant})\mapsto r^{k}}$
(although the relationships between the shift constants for $B_{k}$, $P_{k}$
and $r^{k}$ are rather cumbersome) so that we eventually obtain
\begin{equation}
A_{i}(r^{i})=r^{i}\,,\qquad B_{i}(r^{i})=\zeta _{i}r^{i}\,,\quad i=1,2,3,
\label{AB}
\end{equation}%
where
\begin{equation}
\zeta _{1}=\frac{\xi _{3}\epsilon _{12}-\xi _{2}\epsilon _{13}}{\epsilon
_{12}-\epsilon _{13}}\,,\quad \zeta _{2}=\frac{\xi _{1}\epsilon _{23}-\xi
_{3}\epsilon _{12}}{\epsilon _{23}-\epsilon _{12}}\,,\quad \zeta _{3}=\frac{%
\xi _{1}\epsilon _{23}-\xi _{2}\epsilon _{13}}{\epsilon _{23}-\epsilon _{13}}%
\,,  \label{zeta}
\end{equation}%
\begin{equation}
P_{1}=\frac{\xi _{2}-\xi _{3}}{\epsilon _{12}-\epsilon _{13}}r^{1}+\epsilon
_{23}\,,\quad P_{2}=\frac{\xi _{1}-\xi _{3}}{\epsilon _{12}-\epsilon _{23}}%
r^{2}+\epsilon _{13}\,,\quad P_{3}=\frac{\xi _{1}-\xi _{2}}{\epsilon
_{13}-\epsilon _{23}}r^{3}+\epsilon _{12}\,.  \label{P}
\end{equation}%
Direct verification shows that the diagonal system (\ref{R3}), (\ref{r3}), (%
\ref{AB}), (\ref{zeta}) is indeed equivalent, via (\ref{u3}), (\ref{det3}), (%
\ref{P}), to the original set of conservation laws (\ref{cont}), (\ref{alg}%
), where $v^{k}(\mathbf{u(r)})=V^{k}(\mathbf{r})$.

Thus, system (\ref{cont}), (\ref{alg}) is consistent with formulae (\ref{r3}%
), (\ref{u3}) defined by the St\"{a}ckel matrix (\ref{st}). Therefore, by
Corollary 4.1, the three-component hydrodynamic reduction (\ref{cont}), (\ref%
{alg}) is a linearly degenerate semi-Hamiltonian (i.e. integrable)
hydrodynamic type system.

\textbf{Remark.} As we have seen, the outlined construction has some
additional inherent ``degrees of freedom'', namely, three arbitrary
constants due to non-uniqueness of the St\"ackel matrix specifying a given
linearly degenerate semi-Hamiltonian system. The full set of arbitrary
constants removable by an appropriate change of the Riemann invariants will
appear later in Section 5 where we shall consider $N$-component hydrodynamic
reductions with arbitrary $N \ge 3$.

Using (\ref{r3})--(\ref{P}) we obtain explicit expressions for the
characteristic velocities $V^{k}$ and densities $u^{k}$ in terms of Riemann
invariants,
\begin{eqnarray}
V^{1} &=&\frac{\zeta _{2}r^{2}-\zeta _{3}r^{3}}{r^{2}-r^{3}}\text{, \ \ }
V^{2}=\frac{\zeta _{3}r^{3}-\zeta _{1}r^{1}}{r^{3}-r^{1}}\text{, \ \ }V^{3}=%
\frac{\zeta _{1}r^{1}-\zeta _{2}r^{2}}{r^{1}-r^{2}},  \label{v123} \\
&&  \notag \\
u^{1} &=&P_{1}\frac{r^{2}-r^{3}}{\det \Delta }\text{, \ \ \ \ }u^{2}=P_{2}
\frac{r^{3}-r^{1}}{\det \Delta }\text{, \ \ \ \ }u^{3}=P_{3}\frac{%
r^{1}-r^{2} }{\det \Delta },  \label{u123}
\end{eqnarray}
where
\begin{equation}
\det \Delta =(\zeta _{1}-\zeta _{2})r^{1}r^{2}+(\zeta _{2}-\zeta
_{3})r^{2}r^{3}+(\zeta _{3}-\zeta _{1})r^{3}r^{1}\,.  \label{dec}
\end{equation}
We note that, unlike in the case $N=2$, algebraic system (\ref{alg}) cannot
be resolved for $u^{k}$ in terms of $v^{n}$ for any odd $N$ (cf.
corresponding formulae in Section 2), because determinant of the matrix $%
\mathbf{\hat{A}}$ of linear system (\ref{alg})
\begin{equation*}
\mathbf{\hat{A}u}=\mathbf{b,}
\end{equation*}
where $A_{ik}=\epsilon _{ik}(v^{k}-v^{i})$ and $b_{i}=v^{i}-\xi _{i}$,
equals zero due to its skewsymmetry. For instance, for $N=3$, the
consistency condition of this linear system (i.e. the condition that the
rank of the augmented matrix equals 2) is given by the relation
\begin{equation}
\epsilon _{23}(v^{3}-v^{2})(\xi _{1}-v^{1})+\epsilon _{12}(v^{2}-v^{1})(\xi
_{3}-v^{3})+\epsilon _{13}(v^{1}-v^{3})(\xi _{2}-v^{2})=0.  \label{relv}
\end{equation}%
Direct substitution of $v^j=V^j(\mathbf{r})$ (\ref{v123}) into (\ref{relv})
shows that it satisfies identically.

Using (\ref{u123}), (\ref{dec}), (\ref{zeta}), (\ref{P}) one can express the
Riemann invariants in terms of the densities $u^{k}$ explicitly,
\begin{eqnarray}
r^{1} &=&\frac{(\epsilon _{12}-\epsilon _{13})(\epsilon _{12}\epsilon
_{13}u^{1}+\epsilon _{12}\epsilon _{23}u^{2}+\epsilon _{13}\epsilon
_{23}u^{3}+\epsilon _{23})}{[(\xi _{3}-\xi _{1})\epsilon _{12}+(\xi _{1}-\xi
_{2})\epsilon _{13}]u^{1}-(\xi _{2}-\xi _{3})(\epsilon _{12}u^{2}+\epsilon
_{13}u^{3}+1)},  \notag \\
&&  \notag \\
r^{2} &=&\frac{(\epsilon _{23}-\epsilon _{12})(\epsilon _{12}\epsilon
_{13}u^{1}+\epsilon _{12}\epsilon _{23}u^{2}+\epsilon _{13}\epsilon
_{23}u^{3}+\epsilon _{13})}{[(\xi _{1}-\xi _{2})\epsilon _{23}+(\xi _{2}-\xi
_{3})\epsilon _{12}]u^{2}-(\xi _{3}-\xi _{1})(\epsilon _{12}u^{1}+\epsilon
_{23}u^{3}+1)},  \label{ru} \\
&&  \notag \\
r^{3} &=&\frac{(\epsilon _{13}-\epsilon _{23})(\epsilon _{12}\epsilon
_{13}u^{1}+\epsilon _{12}\epsilon _{23}u^{2}+\epsilon _{13}\epsilon
_{23}u^{3}+\epsilon _{12})}{[(\xi _{2}-\xi _{3})\epsilon _{13}+(\xi _{3}-\xi
_{1})\epsilon _{23}]u^{3}-(\xi _{1}-\xi _{2})(\epsilon _{13}u^{1}+\epsilon
_{23}u^{2}+1)}.  \notag
\end{eqnarray}%
Direct substitution shows that expressions (\ref{ru}) and (\ref{v123}) are
consistent with original algebraic system (\ref{alg}) where $v^{j}=V^{j}(%
\mathbf{r(u)})$.

It is instructive to look at what happens to the diagonal system (\ref{R3})
when the density of one of the components in conservation laws (\ref{cont}),
say $u^{3}$, vanishes. One can see from (\ref{ru}) that if $u^{3}=0$ (this
corresponds to vanishing of $P_{3}$ in (\ref{u123})) then the Riemann
invariant $r^{3}$ becomes a constant,
\begin{equation*}
u^{3}=0:\qquad r^{3}=-\frac{(\epsilon _{23}-\epsilon _{13})\epsilon _{12}}{%
\xi _{1}-\xi _{2}}\equiv r_{0}^{3}\,,
\end{equation*}%
so that the equation for $r^{3}$ satisfies identically and system (\ref{R3})
reduces to its 2-component counterpart (\ref{chap}) for
\begin{equation*}
v^{1}(u^{1},u^{2})=V^{1}(r^2(u^{1},u^{2},0))\,, \qquad
v^{2}(u^{1},u^{2})=V^{2}(r^1(u^{1},u^{2},0))\, ,
\end{equation*}%
as one should expect. Similar reductions occur for $u^{1}=0$ and $u^{2}=0$,
which lead to $r^{1}=r_{0}^{1}=\hbox{constant}$ and $r^{2}=r_{0}^{2}=%
\hbox{constant}$ respectively. As a matter of fact, any function $%
R^{j}(r^{j})$ is also a Riemann invariant so one can choose a new set of
Riemann invariants say $R^{j}=r^{j}-r_{0}^{j}$ so that $R^{j}=0$ when $%
u^{j}=0$. This normalisation could be useful for applications.

Now we consider some special families of solutions to linearly
degenerate system (\ref{R3}), (\ref{v123}).

\medskip

\textit{a) \ Similarity solutions}

One can see that, owing to homogeneity of the characteristic velocities (\ref%
{v123}) as functions of Riemann invariants, system (\ref{R3}) admits
similarity solutions of the form
\begin{equation}  \label{sim}
r^i=\frac{1}{t^{\alpha}}l^i\left(\frac{x}{t}\right)\, , \quad i=1,2,3,
\end{equation}
where $\alpha$ is an arbitrary positive real number and the functions $%
l^i(\tau)$, where $\tau=x/t$, satisfy the system of ordinary differential
equations
\begin{equation}  \label{ode}
(V^i(\mathbf{l})+\tau)\frac{dl^i}{d \tau}+\alpha l^i=0 \, , \quad i=1,2,3.
\end{equation}
Here the functions $V^i(\mathbf{l})$ are obtained from (\ref{v123}) by
replacing $r^i$ with $l^i$. It is not difficult to see that, due to the
structure of the characteristic velocities, the case $\alpha=0$ implies only
a constant solution $l^i= l^i_0$, where $l_0^1, l_0^2, l_0^3$ are arbitrary
constants. If $\alpha \ne 0$, the general solution of (\ref{ode}) can be
found in an implicit form using the generalised hodograph formulae (\ref%
{theo}), where for $N=3$ we substitute, according to (\ref{st}), (\ref{AB}),
$\phi _{k}^{1}(\xi ) \equiv B_k(\xi)= \zeta_k \xi $, $\phi _{k}^{2}(\xi )
\equiv A_k(\xi)=\xi$. To obtain similarity solutions (\ref{sim}) one should
use in (\ref{theo}) $f_i(\xi)= \xi^\beta/c_i$, where $\beta=2+1/\alpha$ and $%
c_i$, $i=1,2,3$, are arbitrary nonzero constants. Then the requirement that
the functions $l^i$ must depend on $\tau = x/t$ alone leads to the algebraic
system
\begin{eqnarray}
\tau &=& c_1 \zeta_1 (l^1)^\gamma + c_2 \zeta_2(l^2)^\gamma + c_3 \zeta_3
(l^3)^\gamma \, ,  \notag \\
&&  \notag \\
-1 &=& c_1(l^1)^\gamma + c_2 (l^2)^\gamma + c_3(l^3)^\gamma \, ,
\label{simsol} \\
&&  \notag \\
0 &=& c_1(l^1)^{\gamma-1} + c_2(l^2)^{\gamma - 1} + c_3(l^3)^{\gamma - 1} \,
,  \notag
\end{eqnarray}
where $\gamma =-1/\alpha$ and we have also replaced $c_i/\gamma \mapsto c_i$%
. Direct substitution shows that solution $l^i$ defined by (\ref{simsol})
indeed satisfies system (\ref{ode}). We note that this family of solutions
is unique to the case $N=3$ and generally does not exist for $N>3$. \medskip

\textit{b) \ Quasiperiodic solutions}

Another interesting type of solutions arises when one introduces in (\ref%
{theo}) (for $N=3$)
\begin{equation*}
f_{1}(\xi )=f_{2}(\xi )=f_{3}(\xi )=\sqrt{R_{7}(\xi )}\,,\qquad R_{7}(\xi )=%
\overset{7}{\underset{n=1}{\prod }}(\xi -E_{n})\,,
\end{equation*}%
where $E_{1}<E_{2}<\dots <E_{7}$ are real constants. Then, according to (\ref%
{AB}), solution (\ref{theo}) assumes the form
\begin{eqnarray}
x &=&\zeta _{1}\overset{r^{1}}{\int }\frac{\xi d\xi }{\sqrt{R_{7}(\xi )}}%
+\zeta _{2}\overset{r^{2}}{\int }\frac{\xi d\xi }{\sqrt{R_{7}(\xi )}}+\zeta
_{3}\overset{r^{3}}{\int }\frac{\xi d\xi }{\sqrt{R_{7}(\xi )}},  \label{1} \\
&&  \notag \\
-t &=&\overset{r^{1}}{\int }\frac{\xi d\xi }{\sqrt{R_{7}(\xi )}}+\overset{%
r^{2}}{\int }\frac{\xi d\xi }{\sqrt{R_{7}(\xi )}}+\overset{r^{3}}{\int }%
\frac{\xi d\xi }{\sqrt{R_{7}(\xi )}},  \label{2} \\
&&  \notag \\
0 &=&\overset{r^{1}}{\int }\frac{d\xi }{\sqrt{R_{7}(\xi )}}+\overset{r^{2}}{%
\int }\frac{d\xi }{\sqrt{R_{7}(\xi )}}+\overset{r^{3}}{\int }\frac{d\xi }{%
\sqrt{R_{7}(\xi )}},  \label{3}
\end{eqnarray}%
which resembles the celebrated system for the multi-gap (here -- three-gap)
solutions of the KdV equation. Unlike (\ref{1}) - (\ref{3}), however, the
three-gap KdV solutions correspond to the St\"{a}ckel matrix (\ref{st}) with
the rows $A_{k}(\xi )=\xi $, $B_{k}(\xi )=\xi ^{2}$ , $k=1,2,3$ \cite{fer91}.

\medskip

\textbf{Proposition 5.1.} \textit{For any constants $\zeta _{1} \ne \zeta
_{2} \ne \zeta _{3} \ne 0$ there exists at least one set $\{E_1, \dots ,
E_6\}$ such that the solution $r^i(x,t)$, $i=1,2,3$ described by (\ref{1}) -
(\ref{3}) is quasi-periodic in $x$ and possibly in $t$.}

\medskip We present here a sketch of the proof. Availability of the solution
in the form (\ref{1}) - (\ref{3}) implies the existence of separate dynamics
of $r^j$-s with respect to $x$ and $t$. Indeed, differentiating (\ref{1}) - (%
\ref{3}) with respect to $x$ for fixed $t$ one readily obtains
\begin{equation}  \label{separx}
\frac{\partial r^{i}}{\partial x}=(r^{j}-r^{k})\frac{\sqrt{R_{7}(r_{i})}}{
\Pi }\, , \qquad i,j,k=1,2,3, \quad i\ne j \ne k\, ,
\end{equation}
where
\begin{equation}  \label{separt0}
\Pi(r_1,r_2,r_3)=(\zeta _{1}-\zeta _{2})r^{1}r^{2}+(\zeta _{2}-\zeta
_{3})r^{2}r^{3}+(\zeta _{3}-\zeta _{1})r^{3}r^{1}=\det \Delta
\end{equation}
-- see (\ref{dec}).

Analogously, differentiating (\ref{1}) - (\ref{3}) with respect to $t$ for
fixed $x$ one obtains
\begin{equation}  \label{separt1}
\frac{\partial r^{i}}{\partial t}=(\zeta _{j}r^{j}-\zeta _{k}r^{k})\frac{%
\sqrt{R_{7}(r_{i})}}{\Pi} \, , \qquad i,j,k=1,2,3, \quad i\ne j \ne k\, .
\end{equation}

One can see that the flows (\ref{separx}) and (\ref{separt1}) are consistent
with the spatio-temporal dynamics (\ref{rim}), (\ref{v123}). We also note
that equations (\ref{separx}), (\ref{separt1}) resemble Dubrovin's equations
for the auxiliary spectrum dynamics in the KdV finite-gap integration
problem (see, for instance, \cite{novTS}).

Let us now suppose that
\begin{equation}  \label{real}
r^{1}\in \lbrack E_{1},E_{2}],\;r^{2}\in \lbrack E_{3},E_{4}],\;r^{3}\in
\lbrack E_{5},E_{6}],
\end{equation}%
so that all $\sqrt{R_{7}(r^{i})}$ are real. The above condition (\ref{real})
means that the point $p=(r^{1},r^{2},r^{3})\in {\mathbb{R}}^{3}$ lies within
the rectangular box $K_{ijk} \in {\mathbb{R}}^{3}$ with the vertices at $%
(E_i, E_j, E_k)$, $i,j,k=1, \dots, 6$, $i\ne j \ne k$.

Now, for any set of the constants $\zeta _{1},\zeta _{2},\zeta _{3} $ there
exists at least one box $K_{i,j,k}=K^{\ast } \in {\mathbb{R}}^{3}$, which is
not intersected by the cone $\Pi (r^{1},r^{2},r^{3})=0$. That is, inside $%
K^{\ast }$ the denominator $\Pi(r_1,r_2,r_3)$ in (\ref{separx}) never
vanishes.

Assume now that the `initial' values of $r^{1},r^{2},r^{3}$ for some $x=x_0$
belong to $K^{\ast }$. Then it follows from (\ref{separx}) that, under the $%
x $-flow ($t=\hbox{const}$), the point $p$ remains inside $K^{\ast }$ and
undergoes \textquotedblleft elastic\textquotedblright\ reflections at the
faces of $K^{\ast }$ as $x$ varies (note that, since $r^j \ne r^k$ for $j
\ne k$, the factor $(r^j - r^k)$ in (\ref{separx}) never vanishes so the
reflections occur only at the faces of $K^*$). Therefore, the motion is
quasi-periodic with respect to $x$ as long as conditions (\ref{real}) are
satisfied. Indeed, the system (\ref{separx}) possesses two integrals (\ref{2}%
) and (\ref{3}) outside the ``resonant'' points, where $\Pi =0$, so it
specifies a quasi-periodic motion on a 3-torus provided conditions (\ref%
{real}) are satisfied. Of course, if conditions (\ref{real}) are not
satisfied at $x=x_0$ the solutions $r^{i}(x)$ may blow up and not be
quasi-periodic.

The proof of quasi-periodicity of the $t$-flow is similar, however, there is
an additional requirement that the factor $(\zeta _{j}r^{j}-\zeta _{k}r^{k})$
in (\ref{separt0}) should not vanish for all $\mathbf{r} \in K^*$ which
might impose additional restrictions on the choice of $E_i$ (that is for
some $\{ E_j\}$ the motion can be quasi-periodic in $x$ but not in $t$).

We note that the quasi-periodicity of the $x$- and $t$-flows can be proved
directly from the solution (\ref{1}) -- (\ref{3}), however the outlined
proof using the dynamical systems arguments is qualitatively more
transparent and more readily yields the \textquotedblleft
resonant\textquotedblright\ restrictions for $x$- and $t$-flows. We also
note that the quasiperiodic solutions could be constructed for $N>3$ as well
(see Section 8.2).

\section{Integrability of $N$-component hydrodynamic reductions}

We now prove our main Theorem 3.1 stating that the $N$-component `cold-gas'
hydrodynamic reduction (\ref{cont}), (\ref{alg}) represents a
semi-Hamiltonian (i.e. integrable) linearly degenerate hydrodynamic type
system. For that, according to Corollary 3.1, it is sufficient to show that
the conservation law densities $u^{i}$ and the transport velocities $v^{i}$
admit parametric representations (\ref{FerSbs}) and (\ref{vstak}), $u^i=U^i({%
\mathbf{r}})$ and $v^{i}(\mathbf{U(r)})=V^{i}(\mathbf{r})$, via $N$
functions $r^k$ in terms of the St\"ackel matrix (\ref{stak}).

We suppose that hydrodynamic type system (\ref{cont}), (\ref{alg}) can be
rewritten in a diagonal form (\ref{rim}), and, moreover, the characteristic
velocities $V^{i}(\mathbf{r})$ coincide with the expressions $v^{i}(\mathbf{%
U }(\mathbf{r}))$.

Now, substitution of (\ref{vstak}), (\ref{FerSbs}) into (\ref{alg}) leads to
the algebraic system%
\begin{equation}
\sum_{k=1}^{N}\epsilon _{ik}(-1)^{k}P_{k}\det \Delta _{ik}^{(12)}=\det
\Delta _{i}^{(2)}-\xi _{i}\det \Delta _{i}^{(1)}, \qquad i=1, \dots, N \, ,
\label{bra}
\end{equation}%
for $P_{k}(r^{k})$ and $\phi _{k}^{i}(r^{k})$. Here the matrix $\Delta
_{ik}^{(12)}$ is the matrix $\Delta $ with \textit{first two} rows and $i$th
and $k$th columns deleted. In the derivation of (\ref{bra}) we have used the
\textit{determinant Sylvester identity} (see, for instance, Gantmacher 1959)
\begin{equation*}
\det \Delta _{ik}^{(12)}=\frac{\det \Delta _{k}^{(1)}\det \Delta
_{i}^{(2)}-\det \Delta _{i}^{(1)}\det \Delta _{k}^{(2)}}{\det \Delta }\, .
\end{equation*}
Expanding the determinants,
\begin{equation*}
\det \Delta _{i}^{(1)}=\sum_{k=1}^{N}\left[ (-1)^{k+1}\phi _{k}^{2}\det
\Delta _{ik}^{(12)}\right] \text{, \ \ \ \ \ \ }\det \Delta
_{i}^{(2)}=\sum_{k=1}^{N}\left[ (-1)^{k+1}\phi _{k}^{1}\det \Delta
_{ik}^{(12)}\right] ,
\end{equation*}%
we rewrite equations (\ref{bra}) as $N$ \textit{nonlinear} systems for $\phi
_{k}^{n}$ \textit{and }$P_{k}$, where $k, n=1, \dots, N$,
\begin{equation}  \label{set}
\sum_{k=1}^{N}(-1)^{k}(\phi _{k}^{1}-\xi _{i}\phi _{k}^{2}+\epsilon
_{ik}P_{k})\det \Delta _{ik}^{(12)}=0, \qquad i=1, \dots, N.
\end{equation}
We recall that $\phi _{k}^{N-1}=r^k$, $\phi _{k}^{N}=1$.

One can now introduce $N$ matrices $\delta _{i}$ obtained from the matrix $%
\Delta $ by deleting the first two rows and the $i$-th column, and adding
the first row with the elements $\phi _{k}^{1}-\xi _{i}\phi
_{k}^{2}+\epsilon _{ik}P_{k}$. Thus, each matrix $\delta_i$ has dimension $%
(N-1) \times (N-1)$. Then the above set of equations (\ref{set}) can be
rewritten as
\begin{equation}  \label{detdelta}
\det \delta _{i}=0, \qquad i=1, \dots, N,
\end{equation}%
which implies that the rows of each of the matrices $\delta _{i}$ must be %
%
\textit{linearly dependent}:
\begin{equation}  \label{theorem}
\begin{split}
C_{i,1}\left( \epsilon _{ik}P_{k} +\phi _{k}^{1}-\xi _{i}\phi
_{k}^{2}\right) +\sum_{n=3}^{N-2}C_{i,n-1}\phi _{k}^{n}&
=C_{i,N-2}r^{k}+C_{i,N-1}, \\
k& =1,\dots N,\ \ i=1,\dots ,N-1\,,\quad k\neq i,
\end{split}%
\end{equation}
where $C_{i,k}$ are arbitrary constants. These conditions can be considered
as $N$ linear systems, for fixed $k$ each. Since all these systems are
consistent the functions $\phi _{k}^{i}$ and $P_{k}$ can be found by solving
system (\ref{theorem}).

Constants $C_{i,1}$ cannot be equal to zero since in that case, according to
(\ref{vstak}), the velocities $V^{i}$ would become undetermined. Therefore,
without loss of generality we can set $C_{i,1}=1$ and the number of free
constants becomes $N(N-2)$. Thus, the following Proposition is valid:
\medskip

\textbf{Proposition 6.1}: \textit{General solution of system (\ref{set}) is
determined by solutions
\begin{equation}  \label{theorem52}
\phi_k^i = \frac{\det{\tilde B_k^i} r^k +\det{\bar B_k^i}}{\det{B_k}},
\qquad P_k = \frac{\det{B_k^{{\small (P)}}}}{\det{B_k}}
\end{equation}
of $N$ linear systems (\ref{theorem}), where $B_k$, $\bar B_{k}^i$, $\tilde
B_k^i$ and $B_k^{{\small (P)}}$ are matrices with elements}
\begin{eqnarray}  \label{array}
&&\tilde b_{kl}^{im} = \bar b_{kl}^{im} = b_{kl}^{{\small (P)}m}=b_{kl}^m =
\left\{
\begin{array}{cc}
1 & \text{for } l = 2 \\
-\xi_l & \text{for } l = 3 \\
C_{m,l-2} & \text{for } l>3%
\end{array}
\right. \text{ if }
\begin{array}{ll}
l \neq i+1 &  \\
l \neq 1 &
\end{array}
\\
&&\text{ and } b_{k\,1}^{im} = \tilde b_{k\,1}^{im} = \bar b_{k\,1}^{im} =
\epsilon_{mk}, \bar b_{ki}^{i+1\;m} = C_{i,N-1}, \ \tilde b_{ki}^{i+1\;m} =
C_{i, N-2}, b_{k\,1}^{{\small (P)}m}= C_{i, N-2} r^k + C_{i,N-1},  \notag
\end{eqnarray}
\textit{where $C_{m,l}$ are arbitrary constants such that $\det B_k \neq 0$.}

\medskip \textbf{Remark:} The set of constants $C_{l,m}$ for which $\det B_k
= 0$ has Lebesque measure zero or requires a very special choice of the
parameters $\eta_k$. The exceptional case is the following: the vectors $%
\mathbf{\xi}$, $\mathbf{1}$ and $\mathbf{\ \epsilon}_k$ are linearly
dependent which yields, according to the definition (\ref{rel1}), a set of
equations for the special values $\eta_k$.

\medskip

Thus, we have proved that all elements of the St\"{a}ckel matrix (\ref{stak}%
) depend linearly on Riemann invariants 
and these elements are determined from the algebraic system (\ref{alg}) up
to $N(N-2)$ arbitrary constants removable by an appropriate change of the
Riemann invariants (for instance, by a shift in the case $N=3$). By
Corollary 4.1, the existence of such a St\"{a}ckel matrix automatically
proves the semi-Hamiltonian and linearly-degenerate properties of the
hydrodynamic reductions (\ref{cont}), (\ref{alg}).

\medskip

Now, our main Theorem 3.1 is proved.

\section{Riemann invariants and characteristic velocities: explicit
construction}

The construction described in Sections 3 and 6 provides a proof of the
existence of Riemann invariants for system (\ref{cont}), (\ref{alg}) for
arbitrary $N$. The Riemann invariants are found to parameterise system (\ref%
{cont}), (\ref{alg}) via the sole St\"{a}ckel matrix, which, by Corollary
4.1, implies linear degeneracy and integrability of this system. Explicit
representations for conservation law densities $u^{i}$ and transport
velocities $v^{j}$ in terms of the Riemann invariants are given by
Ferapontov \cite{fer91} formulae (\ref{FerSbs}), (\ref{vstak}) where the
entries $\phi _{k}^{n}$ of the St\"{a}ckel matrix (\ref{stak}) and the
functions $P_{k}(r^{k})$ are defined by formulae (\ref{theorem52}) -- (\ref%
{array}). Using the functions $\phi _{k}^{n}$ one also obtains the
generalised hodograph solutions (\ref{theo}).

The outlined procedure, while providing general theoretical framework for
the study of the `cold-gas' reductions of the kinetic equation for a soliton
gas, seems to be not very convenient from the viewpoint of practical
calculations. It also involves $N(N-2)$ intermediate constants $C_{l,m}$,
which introduce an additional unnecessary complication. It is, thus,
desirable to have more direct representations for the Riemann invariants and
characteristic velocities, which will also be free from these intermediate
arbitrary constants.


We shall make use of the Theorem 3.1 and show that, once the linear
degeneracy and integrability properties of system (\ref{cont}), (\ref{alg})
are established, explicit relations between the Riemann invariants $\mathbf{r%
}$ and the conserved densities $\mathbf{u}$ can be found by a relatively
straightforward calculation. The calculation will involve the properties of
the Lam\'e coefficients outlined in Section 4.

First, without loss of generality we choose the following normalization (see
(\ref{uH}))
\begin{equation}
u^{k}=\bar{H}_{k},  \label{norm}
\end{equation}
where $\bar{H}_{k}$'s are the Lam\'e coefficients (\ref{lam}). Now, using
Theorem 3.1 we assume that hydrodynamic type system (\ref{cont}), (\ref{alg}%
) can be rewritten in a diagonal form (\ref{rim}), so that $u^i=U^i({\mathbf{%
r}})$ and $v^{i}(\mathbf{U(r)})=V^{i}(\mathbf{r})$. For convenience, in what
follows we shall use small $u$'s and $v$'s only, assuming that $u_j=u_j(%
\mathbf{r}) \equiv U^j(\mathbf{r})$, $v_j=v_j(\mathbf{r})\equiv V_j(\mathbf{r%
})$.

To obtain explicit formulae for the Riemann invariants of the hydrodynamic
reduction (\ref{cont}), (\ref{alg}) we need first to prove its so-called
\textquotedblleft Egorov\textquotedblright\ property.

\textbf{Definition 7.1} (Pavlov $\&$ Tsarev 2003 \cite{pavts03}): \textit{%
Semi-Hamiltonian hydrodynamic type system} (\ref{gen}) \textit{is called the
Egorov, if a \underline{sole} pair of conservation laws}%
\begin{equation}
\partial _{t}a(\mathbf{u})=\partial _{x}b(\mathbf{u})\text{, \ \ \ }\partial
_{t}b(\mathbf{u})=\partial _{x}c(\mathbf{u})  \label{Egdef}
\end{equation}%
\textit{exists.}

\medskip It was proved in \cite{pavts03}, that
\begin{equation}
\partial _{i}a=\bar{H}_{i}^{2}\text{, \ \ \ }\partial _{i}b=\tilde{H}_{i}%
\bar{H}_{i}\text{, \ \ \ }\partial _{i}c=\tilde{H}_{i}^{2},  \label{egor}
\end{equation}%
(see (\ref{lam}) and (\ref{sec}) for the definitions of $\bar {H}_i$ and $%
\tilde{H}_i$) while the corresponding rotation coefficients (\ref{rot})
become symmetric, i.e.
\begin{equation}
\beta _{ik}=\beta _{ki},\text{ \ }i\neq k.  \label{syma}
\end{equation}

Another important fact proven in \cite{pavts03} is that all commuting flows
to a semi-Hamiltonian Egorov system are also Egorov so commuting flow (\ref%
{comm}) possesses a similar pair of conservation laws%
\begin{equation*}
\partial _{\tau }a(\mathbf{u})=\partial _{x}h(\mathbf{u})\text{, \ \ \ }%
\partial _{\tau }h(\mathbf{u})=\partial _{x}g(\mathbf{u}),
\end{equation*}%
where%
\begin{equation}
\partial _{i}h=H_{i}\bar{H}_{i}\text{, \ \ \ }\partial _{i}g=H_{i}^{2}.
\label{poten}
\end{equation}

Now we prove the following

\textbf{Lemma 7.1}: Hydrodynamic reductions (\ref{cont}), (\ref{alg}) are
Egorov.

\textbf{Proof}: We consider the sum of conservation laws (\ref{cont}), (\ref%
{alg})
\begin{equation}  \label{cons1}
\partial _{t}\left( \sum u^{k}\right) =\partial _{x}\left( \sum
u^{k}v^{k}\right) =\partial _{x}\left[ \underset{k=1}{\overset{N}{\sum }}%
u^{k}\left( \xi _{k}+\sum_{m\neq k}\epsilon _{km}u^{m}(v^{m}-v^{k})\right) %
\right]
\end{equation}%
One can see that, since the matrix $\epsilon _{ik}$ is symmetric, the last
term in r.h.s. of (\ref{cons1}) vanishes. Thus, (\ref{cons1}) simplifies to
the form%
\begin{equation}  \label{cons2}
\partial _{t}\left( \sum u^{k}\right) =\partial _{x}\left( \sum \xi
_{k}u^{k}\right) .
\end{equation}%
However, the \textit{flux} $\Sigma \xi _{k}u^{k}$ of conservation law (\ref%
{cons2}) is nothing but the \textit{density} of another conservation law
which can be obtained by the same summation but with the special weights $%
\xi _{i}$, i.e.%
\begin{equation*}
\partial _{t}\left( \sum \xi _{k}u^{k}\right) =\partial _{x}\left( \sum \xi
_{k}u^{k}v^{k}\right) .
\end{equation*}%
Comparison with definition (\ref{Egdef}) implies that in our case
\begin{equation}
a=\sum u^{m},\text{ \ \ \ }b=\sum \xi _{m}u^{m}\equiv \sum u^{m}v^{m},\text{
\ \ \ }c=\sum \xi _{m}u^{m}v^{m},  \label{pots}
\end{equation}
which completes the proof.

Now we formulate the following

\textbf{Theorem 7.1}: \textit{The Riemann invariants of $N$-component
hydrodynamic reductions (\ref{cont}), (\ref{alg}) can be found explicitly as}
\begin{equation}  \label{inv0}
r^{i}=-\frac{1}{u^{i}}\left( 1+\underset{m\neq i}{\sum }\epsilon
_{im}u^{m}\right) \, , \qquad i=1, \dots, N.
\end{equation}

\textbf{Proof}:

For the sake of completeness of our construction we first show that the
linear degeneracy property (\ref{lin}) of system (\ref{cont}), (\ref{alg})
readily follows from the (already established) existence of the Riemann invariants $%
r^{k} $. Indeed, differentiating (\ref{alg}) with respect to the Riemann
invariant $r^{i}$ and taking into account that (see (\ref{lam}), (\ref{norm}%
))
\begin{equation*}
\partial _{i}\ln u^{k}=\frac{\partial _{i}v^{k}}{v^{i}-v^{k}}\text{, \ }%
i\neq k,
\end{equation*}%
we obtain the expression
\begin{equation*}
\partial _{i}v^{i}=\underset{m\neq i}{\sum }\epsilon
_{im}(v^{m}-v^{i})\partial _{i}u^{m}+\underset{m\neq i}{\sum }\epsilon
_{im}u^{m}(\partial _{i}v^{m}-\partial _{i}v^{i})\,,
\end{equation*}%
which reduces, on using (\ref{den}), to the form
\begin{equation}
\partial _{i}v^{i}\left( 1+\underset{m\neq i}{\sum }\epsilon
_{im}u^{m}\right) =0\,.  \label{exp3}
\end{equation}%
Equation (\ref{exp3}) can only be satisfied if $\partial _{i}v^{i}=0$ for
all $i$ (otherwise the field variables $u^{m}$ in the algebraic system (\ref%
{alg}) would cease to be independent). Thus system (\ref{cont}), (\ref{alg})
is indeed linearly degenerate.

Now, differentiation of algebraic system (\ref{alg}) with respect to the
Riemann invariant $r^{k}$ yields
\begin{equation*}
\partial _{k}v^{i}=\underset{m\neq i,k}{\sum }\epsilon _{im}u^{m}(\partial
_{k}v^{m}-\partial _{k}v^{i})+\underset{m\neq i,k}{\sum }\epsilon
_{im}(v^{m}-v^{i})\partial _{k}u^{m}+\epsilon _{ik}u^{k}(\partial
_{k}v^{k}-\partial _{k}v^{i})+\epsilon _{ik}(v^{k}-v^{i})\partial _{k}u^{k},
\end{equation*}%
which reduces, with an account of (\ref{den}) and the linear degeneracy
property, to
\begin{equation*}
(v^{k}-v^{i})\left[ \left( 1+\underset{m\neq i}{\sum }\epsilon
_{im}u^{m}\right) \partial _{k}\ln u^{i}-\underset{m\neq i}{\sum }\epsilon
_{im}\partial _{k}u^{m}\right] =0.
\end{equation*}%
Since all characteristic velocities $v^{k}$ are distinct, the expression in
square brackets must vanish for any pair of indices $i$ and $k$, i.e. we
have
\begin{equation}
\partial _{k}\ln u^{i}=\frac{\underset{m\neq i}{\sum }\epsilon _{im}\partial
_{k}u^{m}}{1+\overset{}{\underset{m\neq i}{\sum }}\epsilon _{im}u^{m}},\text{
\ \ }k\neq i\,.  \label{exp1}
\end{equation}%
Integration of (\ref{exp1}) yields
\begin{equation}
\underset{m\neq i}{\sum }\epsilon _{im}u^{m}+R_{i}(r^{i})u^{i}=-1,
\label{lame}
\end{equation}%
where $R_{i}(r^{i})$, $i=1,\dots ,N$ are arbitrary functions.

\medskip

We now differentiate (\ref{lame}) with respect to the Riemann invariants $%
r^{i}$ and $r^{k}$, which gives, on using (\ref{norm}) and (\ref{rot}),
\begin{equation}
\underset{m\neq i}{\sum }\epsilon _{im}\beta _{im}+R_{i}^{\prime
}(r^{i})+R_{i}(r^{i})\partial _{i}\ln \bar{H}_{i}=0  \label{dri}
\end{equation}%
and
\begin{equation}
\underset{m\neq i,k}{\sum }\epsilon _{im}\beta _{km}+R_{i}(r^{i})\beta
_{ki}+\epsilon _{ik}\partial _{k}\ln \bar{H}_{k}=0  \label{drk}
\end{equation}%
respectively. Substitution of (\ref{pots}) into (\ref{egor}) gives
\begin{equation}
\bar{H}_{i}=\underset{m\neq i}{\sum }\beta _{im}+\partial _{i}\ln \bar{H}_{i}%
\text{, \ \ \ }\tilde{H}_{i}=\xi _{i}\bar{H}_{i}+\underset{m\neq i}{\sum }%
(\xi _{m}-\xi _{i})\beta _{im}.  \label{com}
\end{equation}%
By expressing $\partial _{i}\ln \bar{H}_{i}$ from the above first equation, (%
\ref{dri}) and (\ref{drk}) reduce to the form%
\begin{eqnarray}
R_{i}(r^{i})\bar{H}_{i} =R_{i}(r^{i})\underset{m\neq i}{\sum }\beta _{im}-%
\underset{m\neq i}{\sum }\epsilon _{im}\beta _{im}-R_{i}^{\prime }(r^{i}),
\notag \\
&&  \label{compa} \\
\epsilon _{im}\bar{H}_{m} =\epsilon _{im}\underset{n\neq m}{\sum }\beta
_{nm}-\underset{n\neq i,m}{\sum }\epsilon _{in}\beta _{nm}-R_{i}(r^{i})\beta
_{im}.  \notag
\end{eqnarray}%
Substitution of the expressions $R_{i}(r^{i})\bar{H}_{i}$ and $\epsilon _{im}%
\bar{H}_{m}$ into (\ref{lame}) yields a set of constraints $R_{i}^{\prime
}(r^{i})=1$, i.e. $R_{i}(r^{i})=r^{i}+\alpha _{i}$, where $\alpha _{i}$ are
arbitrary constants. Since any function of a Riemann invariant is a
Riemann invariant as well one can put without loss of generality that $%
R_{i}(r^{i})=r^{i}$. Then (\ref{lame}) reduces to (\ref{inv0}). The Theorem
is proved.

\medskip

Taking into account $R_{i}(r^{i})=r^{i}$ and eliminating $\bar{H}_{i}$ from (%
\ref{compa}) we arrive at the linear algebraic system%
\begin{equation}
\underset{m\neq i,k}{\sum }(r^{i}\epsilon _{km}-\epsilon _{ik}\epsilon
_{im})\beta _{im}+(r^{i}r^{k}-\epsilon _{ik}^{2})\beta _{ik}=\epsilon _{ik},%
\text{ \ \ }i\neq k  \label{z}
\end{equation}%
for the rotation coefficients $\beta _{ik}$, while (\ref{compa}) reduces
(cf. the first formula in (\ref{com})) to%
\begin{equation}
\bar{H}_{i}=\underset{m\neq i}{\sum }\left( 1-\frac{\epsilon _{im}}{r^{i}}%
\right) \beta _{im}-\frac{1}{r^{i}}.  \label{third}
\end{equation}

Let us introduce a matrix $\boldsymbol{\epsilon }$ such that its
off-diagonal coefficients are the aforementioned symmetric constants $%
\epsilon _{ik}$, while the diagonal coefficients $\epsilon _{ii}=r^{i}$.

\textbf{Theorem 7.2}: \textit{The rotation coefficients }$\beta _{ik}$
\textit{satisfying linear algebraic system }(\ref{z}) \textit{are the
off-diagonal components of the matrix }\textit{\ inverse to the matrix }$-%
\boldsymbol{\epsilon }$, \textit{i.e.}%
\begin{equation}
\underset{m=1}{\overset{N}{\sum }}\epsilon _{im}\beta _{km}=-\delta _{ik}.
\label{45}
\end{equation}%
%
%
%
%
%
%
%
%
%
%
%
%
%
%

\textbf{Proof}: We introduce the functions $\beta _{ii}(\mathbf{r})$ so that
expression (\ref{55}) could be extended to the full set of indices, i.e. we
will have
\begin{equation}
\partial _{i}\beta _{jk}=\beta _{ji}\beta _{ik}\quad \forall i,j,k\,.
\label{555}
\end{equation}%
It is easy to check that (\ref{555}) is valid for \textit{any curvilinear
coordinate system associated with semi-Hamiltonian Egorov linearly
degenerate hydrodynamic type system} (see (\ref{lin}) and (\ref{egor})),
i.e. if and only if the rotation coefficients $\beta _{ik}$ are symmetric
(see (\ref{syma})) and determined by (\ref{ld}), where the functions
\begin{equation}
\beta _{ii}(\mathbf{r})\equiv \partial _{i}\ln \bar{H}_{i}\,.  \label{extra}
\end{equation}%
Indeed, the above set of equations (\ref{555}) for two distinct indices
(just two choices) reduces to the form%
\begin{equation}
\partial _{k}\beta _{jk}=\beta _{jk}\beta _{kk},\text{ \ \ }\partial
_{i}\beta _{kk}=\beta _{ik}^{2}.  \label{dega}
\end{equation}%
The first part of these equations is nothing else but (\ref{ld}) while the
second part is just the well-known property of \textit{any} curvilinear
coordinate net (see \cite{darboux}): the scalar potential $V$ is determined
by its second derivatives, i.e.%
\begin{equation}
\partial _{ik}^{2}V=\beta _{ik}\beta _{ki},\text{ \ \ }k\neq i\,.
\label{pot}
\end{equation}%
Thus, in the Egorov (symmetric) case, the above property (\ref{pot})
simplifies to
\begin{equation}
\partial _{ik}^{2}V=\beta _{ik}^{2},\text{ \ \ }k\neq i\,.  \label{prop1}
\end{equation}%
Comparing this formula and the second formula in (\ref{dega}), one can
conclude that $\beta _{kk}=\partial _{k}V$. If all indices in (\ref{dega})
coincide, the last nontrivial consequence given by%
\begin{equation}
\partial _{k}\frac{1}{\beta _{kk}}=-1  \label{int}
\end{equation}%
allows one to integrate (step-by-step) nonlinear system in partial
derivatives (\ref{555}). Instead of this direct, but somewhat complicated
procedure, we shall use a more sophisticated but technically much more
simple approach to the derivation of general solution of system (\ref{555}%
). First, let us introduce the combinations
\begin{equation}
A_{ik}=\underset{m=1}{\overset{N}{\sum }}\epsilon _{im}\beta _{km}
\label{aik}
\end{equation}%
(we recall that $\epsilon _{ii}=r^{i}$). Then (\ref{z}) reads as follows%
\begin{equation}
r^{i}A_{ik}=\epsilon _{ik}(1+A_{ii}),\text{ \ \ }i\neq k.  \label{44}
\end{equation}%
Differentiation of (\ref{aik}) with respect to the Riemann invariants $%
r^{i},r^{k},r^{j}$ leads to the system%
\begin{equation*}
\partial _{i}A_{ik}=\beta _{ik}(1+A_{ii}),\text{ \ \ }\partial
_{k}A_{ik}=\beta _{kk}A_{ik},\text{ \ \ }\partial _{j}A_{ik}=\beta
_{jk}A_{ij},\text{ \ \ }i\neq k.
\end{equation*}%
Compatibility conditions imply just one extra equation
\begin{equation*}
\partial _{k}A_{ii}=\beta _{ik}A_{ik},\text{ \ \ }i\neq k.
\end{equation*}%
Now we differentiate (\ref{44}) with respect to the Riemann invariant $r^{j}$
to obtain
\begin{equation*}
(r^{i}\beta _{jk}-\epsilon _{ik}\beta _{ij})A_{ij}=0,\text{ \ }i\neq j\neq k.
\end{equation*}%
Since expressions $r^{i}\beta _{jk}-\epsilon _{ik}\beta _{ij}$ cannot vanish
identically, we have the only possible choice: $A_{ik}=0$ for each pair of
distinct indices, and $A_{ii}=-1$ (see (\ref{44})). Thus, we conclude that (%
\ref{aik}) reduces to the form (\ref{45}) (let us emphasize one more time
that $\epsilon _{ii}\equiv r^{i}$, while all the other $\epsilon
_{jk}=\epsilon _{kj}$ are constants). The matrix $\boldsymbol{\epsilon }$
contains $N(N-1)/2$ arbitrary constants $\epsilon _{ik}$, then all
components of the matrix $\boldsymbol{\beta }$ are parameterised by these $%
N(N-1)/2$ arbitrary constants. On the other hand, (\ref{555}) is an
overdetermined system, where \textit{all} first derivatives of $\beta _{ik}$
are expressed via $\beta _{jn}$ only. Thus, a general solution of system in
partial derivatives (\ref{555}) must depend on $N(N+1)/2$ arbitrary
constants, because this system is written for $N(N+1)/2$ functions $\beta
_{ik}$ (these are $N(N-1)/2$ symmetric off-diagonal elements, i.e. rotation
coefficients $\beta _{ik}$; and $N$ diagonal components $\beta _{kk}$). It
means, that the inverse matrix $\boldsymbol{\epsilon }$ contains extra $N$
arbitrary constants $\alpha _{i}$ which are nothing but the shifts of the
Riemann invariants $r^{i}$ located on the diagonal (see the end of the proof
of Theorem 7.1). Then these $N$ shift constants can be removed without loss
of generality. The Theorem is proved.

\medskip In particular, for $N=3$ we have from (\ref{45}) the explicit
expressions for $\beta _{ik}$:
\begin{eqnarray}
\beta _{12} &=&\frac{r^{3}\epsilon _{12}-\epsilon _{13}\epsilon _{23}}{%
r^{1}r^{2}r^{3}-r^{1}\epsilon _{23}^{2}-r^{2}\epsilon
_{13}^{2}-r^{3}\epsilon _{12}^{2}+2\epsilon _{12}\epsilon _{13}\epsilon _{23}%
},  \notag \\
\beta _{13} &=&\frac{r^{2}\epsilon _{13}-\epsilon _{12}\epsilon _{23}}{%
r^{1}r^{2}r^{3}-r^{1}\epsilon _{23}^{2}-r^{2}\epsilon
_{13}^{2}-r^{3}\epsilon _{12}^{2}+2\epsilon _{12}\epsilon _{13}\epsilon _{23}%
},  \label{bet} \\
\beta _{23} &=&\frac{r^{1}\epsilon _{23}-\epsilon _{12}\epsilon _{13}}{%
r^{1}r^{2}r^{3}-r^{1}\epsilon _{23}^{2}-r^{2}\epsilon
_{13}^{2}-r^{3}\epsilon _{12}^{2}+2\epsilon _{12}\epsilon _{13}\epsilon _{23}%
};  \notag
\end{eqnarray}%
\begin{eqnarray*}
\beta _{11} &=&\frac{-r^{2}r^{3}+\epsilon _{23}^{2}}{r^{1}r^{2}r^{3}-r^{1}%
\epsilon _{23}^{2}-r^{2}\epsilon _{13}^{2}-r^{3}\epsilon _{12}^{2}+2\epsilon
_{12}\epsilon _{13}\epsilon _{23}}, \\
\beta _{22} &=&\frac{-r^{1}r^{3}+\epsilon _{13}^{2}}{r^{1}r^{2}r^{3}-r^{1}%
\epsilon _{23}^{2}-r^{2}\epsilon _{13}^{2}-r^{3}\epsilon _{12}^{2}+2\epsilon
_{12}\epsilon _{13}\epsilon _{23}}, \\
\beta _{33} &=&\frac{-r^{1}r^{2}+\epsilon _{12}^{2}}{r^{1}r^{2}r^{3}-r^{1}%
\epsilon _{23}^{2}-r^{2}\epsilon _{13}^{2}-r^{3}\epsilon _{12}^{2}+2\epsilon
_{12}\epsilon _{13}\epsilon _{23}}.
\end{eqnarray*}

\medskip \textbf{Remark.} \ Note that equation (\ref{45}), despite of having
a simpler form than original equation (\ref{z}), is more general as it
defines \textit{all} (not only off-diagonal) components $\beta _{ik}$ in
terms of Riemann invariants. Thus, the rotation coefficients $\beta_{ik}, \
i \ne k$ satisfy both systems (\ref{z}) and (\ref{555}) and are completely
defined in terms of the matrix $\boldsymbol{\epsilon }$.

\medskip

As a by-product of the proof of Theorem 7.2 we obtain the following important

\textbf{Corollary 7.1 } Since system (\ref{555}) describes rotation
coefficients $\beta_{ik}$ associated with hydrodynamic type systems
possessing \textit{simultaneously} Egorov and linear degeneracy properties,
we conclude that our reduction (\ref{cont}), (\ref{alg}) of the kinetic
equation (\ref{kin1}) is \textit{the only} (up to unessential
transformations) hydrodynamic type system possessing both these properties.


\medskip Now, using (\ref{sec}), (\ref{norm}), (\ref{com}), (\ref{third})
and (\ref{extra}) we formulate the main result of this Section:

\textit{Algebraic relations} (\ref{alg}) \textit{can be resolved in a
parametric form in terms of the Riemann invariants:}%
\begin{equation}
u^{i}=\underset{m=1}{\overset{N}{\sum }}\beta _{im},\text{ \ \ \ \ }v^{i}=%
\frac{1}{u^{i}}\underset{m=1}{\overset{N}{\sum }}\xi _{m}\beta _{im},
\label{main}
\end{equation}%
\textit{where the symmetric coefficients} $\beta _{ik}$ \textit{are elements
of the matrix }$-\boldsymbol{\epsilon }^{-1}$ (see (\ref{45})).  As a matter of fact,
the first formula in (\ref{main})  represents the inversion of formula (\ref{inv0}). As one can see,
this inversion is rather nontrivial.

\medskip In particular, for $N=3$ we have from (\ref{main}) the explicit
expressions for conservation law densities $u^{i}$ and characteristic
velocities $v^{k}$%
\begin{eqnarray}
u^{1} &=&\frac{-r^{2}r^{3}+r^{2}\epsilon _{13}+r^{3}\epsilon _{12}-\epsilon
_{12}\epsilon _{23}-\epsilon _{13}\epsilon _{23}+\epsilon _{23}^{2}}{%
r^{1}r^{2}r^{3}-r^{1}\epsilon _{23}^{2}-r^{2}\epsilon
_{13}^{2}-r^{3}\epsilon _{12}^{2}+2\epsilon _{12}\epsilon _{13}\epsilon _{23}%
},  \notag \\
u^{2} &=&\frac{-r^{1}r^{3}+r^{1}\epsilon _{23}+r^{3}\epsilon _{12}-\epsilon
_{12}\epsilon _{13}-\epsilon _{13}\epsilon _{23}+\epsilon _{13}^{2}}{%
r^{1}r^{2}r^{3}-r^{1}\epsilon _{23}^{2}-r^{2}\epsilon
_{13}^{2}-r^{3}\epsilon _{12}^{2}+2\epsilon _{12}\epsilon _{13}\epsilon _{23}%
},  \label{dense} \\
u^{3} &=&\frac{-r^{1}r^{2}+r^{1}\epsilon _{23}+r^{2}\epsilon _{13}-\epsilon
_{12}\epsilon _{13}-\epsilon _{12}\epsilon _{23}+\epsilon _{12}^{2}}{%
r^{1}r^{2}r^{3}-r^{1}\epsilon _{23}^{2}-r^{2}\epsilon
_{13}^{2}-r^{3}\epsilon _{12}^{2}+2\epsilon _{12}\epsilon _{13}\epsilon _{23}%
},  \notag
\end{eqnarray}

\begin{eqnarray}
v^{1} &=&\frac{\xi _{1}\left( \epsilon _{23}^{2}-r^{2}r^{3}\right) +\xi
_{2}\left( r^{3}\epsilon _{12}-\epsilon _{13}\epsilon _{23}\right) +\xi
_{3}\left( r^{2}\epsilon _{13}-\epsilon _{12}\epsilon _{23}\right) }{%
\epsilon _{23}^{2}-r^{2}r^{3}+\epsilon _{13}\left( r^{2}-\epsilon
_{23}\right) +\epsilon _{12}\left( r^{3}-\epsilon _{23}\right) },  \notag \\
v^{2} &=&\frac{\xi _{2}\left( \epsilon _{13}^{2}-r^{1}r^{3}\right) +\xi
_{3}\left( r^{1}\epsilon _{23}-\epsilon _{12}\epsilon _{13}\right) +\xi
_{1}\left( r^{3}\epsilon _{12}-\epsilon _{13}\epsilon _{23}\right) }{%
\epsilon _{13}^{2}-r^{1}r^{3}+\epsilon _{13}(r^{3}-\epsilon _{13})+\epsilon
_{23}(r^{1}-\epsilon _{13})},  \label{explicit3} \\
v^{3} &=&\frac{\xi _{3}\left( \epsilon _{12}^{2}-r^{1}r^{2}\right) +\xi
_{2}\left( r^{1}\epsilon _{23}-\epsilon _{12}\epsilon _{13}\right) +\xi
_{1}\left( r^{2}\epsilon _{13}-\epsilon _{12}\epsilon _{23}\right) }{%
\epsilon _{12}^{2}-r^{1}r^{2}+\epsilon _{13}(r^{2}-\epsilon _{12})+\epsilon
_{23}(r^{1}-\epsilon _{12})}.  \notag
\end{eqnarray}%
One can observe that formulae (\ref{explicit3}) do not coincide with
representation (\ref{v123}), (\ref{zeta}) obtained earlier for the same
family of the characteristic velocities. The reason is that the two
representations correspond to different choices of the Riemann invariants
(we recall one more time that any function of a Riemann invariant is a
Riemann invariant as well). The relationship between these two equivalent
sets of the Riemann invariants can be obtained by equating the
characteristic velocities (\ref{v123}) and (\ref{explicit3}) (or,
alternatively, the densities (\ref{u123}) and (\ref{dense})), which do not
depend on a particular normalization of the Riemann invariants. It is more
convenient, however, to get the sought relationship by a substitution (\ref%
{dense}) into (\ref{ru}), where we replace $r_i$ with $\tilde r_i$. As a
result we get
\begin{eqnarray}
\tilde{r}^{1} &=&\frac{(\epsilon _{13}-\epsilon _{12})(\epsilon
_{23}r^{1}-\epsilon _{12}\epsilon _{13})}{(\xi _{2}-\xi _{3})r^{1}+(\xi
_{3}-\xi _{1})\epsilon _{12}+(\xi _{1}-\xi _{2})\epsilon _{13}},  \notag \\
\tilde{r}^{2} &=&\frac{(\epsilon _{12}-\epsilon _{23})(\epsilon
_{13}r^{2}-\epsilon _{12}\epsilon _{23})}{(\xi _{3}-\xi _{1})r^{2}+(\xi
_{1}-\xi _{2})\epsilon _{23}+(\xi _{2}-\xi _{3})\epsilon _{12}},  \label{rr}
\\
\tilde{r}^{3} &=&\frac{(\epsilon _{23}-\epsilon _{13})(\epsilon
_{12}r^{3}-\epsilon _{13}\epsilon _{23})}{(\xi _{1}-\xi _{2})r^{3}+(\xi
_{2}-\xi _{3})\epsilon _{13}+(\xi _{3}-\xi _{1})\epsilon _{23}}.  \notag
\end{eqnarray}%
Here by $\tilde{r}^{1}$, $\tilde{r}^{2}$, $\tilde{r}^{3}$ we denote the
`old' Riemann invariants as in Section 5.

Note that the choice (\ref{rr}) of the Riemann invariants leads to the
homogeneous expressions (\ref{v123}) for the characteristic velocities,
which makes possible the construction of the similarity solutions (\ref%
{simsol}). Such a possibility, however, is unique to the case $N=3$ as for $%
N>3$ the general rational substitution of the type (\ref{rr}) will not allow
for the elimination of all inhomogeneous terms (unless one has a very
special set of the coefficients $\epsilon_{ij}$, $\xi_k$). As a consequence,
the family of the similarity solutions (\ref{sim}) exists \textit{only} for
the case $N=3$ which makes this case special.

\medskip

\section{Commuting hydrodynamic flows}

\subsection{General explicit representation}

Commuting flows to semi-Hamiltonian linearly degenerate system (\ref{cont}),
(\ref{alg}) are defined in terms of the Riemann invariants by equations (\ref%
{comm}), (\ref{ts}). We recall that, according to Proposition 4.1, only $N-2$
of the commuting flows are linearly degenerate (excluding the `trivial'
flows specified by linear combinations of the constant characteristic
velocity $\mathbf{1}$ and the characteristic velocity $\mathbf{v}$ of the
original flow (\ref{rim})). The general solution of the generalised
hodograph equations (\ref{ts}) specifying commuting flows was obtained by
Ferapontov \cite{fer91} in terms of the St\"ackel matrix entries (see
Theorem 4.2). Here we are interested in a more explicit representation of
the commuting flows for the specific system (\ref{cont}), (\ref{alg}). For
that, instead of integrating system (\ref{ts}), we take advantage of the
fact that our linearly degenerate system (\ref{cont}), (\ref{alg}) is
Egorov. In that case, the commuting flows can be found explicitly.

We first observe that any conservation law density $h$ for linearly
degenerate hydrodynamic type system (\ref{cont}) can be represented in the
form (see (\ref{uH}) or (\ref{FerSbs}))
\begin{equation}  \label{h}
h=\underset{k=1}{\overset{N}{\sum }}u^{k}P_{k}(r^{k}),
\end{equation}
with $N$ arbitrary functions $P_{k}(r^{k})$ of a single variable. Then we
make use of

\textbf{Lemma 8.1} (Pavlov $\&$ Tsarev 2003 \cite{pavts03}): \textit{All
commuting flows (\ref{comm}), (\ref{ts}) in the Egorov case are specified by
the expression (see (\ref{coma}), (\ref{egor}), (\ref{poten}))}%
\begin{equation}
W^{i}=\frac{H_{i}}{\bar{H}_{i}}=\frac{\partial _{i}h}{\partial _{i}a}.
\label{comeg}
\end{equation}

Substituting (\ref{h}), (\ref{pots}) into (\ref{comeg}) and using the first
formula from (\ref{com}) we obtain an explicit representation for the
characteristic velocities of the commuting flows (\ref{comm}),
\begin{equation}
W^{i}=P_{i}(r^{i})+\frac{1}{\bar{H}_{i}}\left( P_{i}^{\prime }(r^{i})+%
\underset{m\neq i}{\sum }(P_{m}(r^{m})-P_{i}(r^{i}))\beta _{im}\right) .
\label{comchar}
\end{equation}%
We recall that $P_{k}(r^{k})$, $k=1,\dots ,N$ are arbitrary functions and
the dependence of the rotation coefficients $\beta _{im}$ on the Riemann
invariants is found by inversion of the matrix $-\boldsymbol{\mathbf{%
\epsilon }}$ (see (\ref{45})). If $P_{k}(r^{k})=1$, (\ref{comchar}) reduces
to $W^{i}=1$; if $P_{k}(r^{k})=\xi _{k}$, it reduces to the second formula
in (\ref{main}), i.e. to hydrodynamic reduction (\ref{cont}), (\ref{alg})
itself.

\subsection{generalised hodograph method}

Taking into account the Combescure transformation (\ref{coma}) and formula (%
\ref{sec}) the generalised hodograph solution (\ref{hod}) can be represented
in a symmetric form
\begin{equation}
x\bar{H}_{i}+t\tilde{H}_{i}=H_{i}(\mathbf{r}).  \label{hodog}
\end{equation}%
Since (see (\ref{main}), (\ref{sec}), (\ref{norm}))%
\begin{equation*}
\bar{H}_{i}=\underset{m=1}{\overset{N}{\sum }}\beta _{im},\text{ \ \ \ \ }%
\tilde{H}_{i}=\underset{m=1}{\overset{N}{\sum }}\xi _{m}\beta _{im},
\end{equation*}%
expression (\ref{hodog}), with an account of (\ref{coma}), (\ref{comchar}),
assumes the form%
\begin{equation}
\underset{m=1}{\overset{N}{\sum }}(x+\xi _{m}t)\beta _{km}=P_{k}^{\prime
}(r^{k})+\underset{m=1}{\overset{N}{\sum }}P_{m}(r^{m})\beta _{km}.
\label{fgh}
\end{equation}%
Multiplying equation (\ref{fgh}) through by the matrix $\boldsymbol{\epsilon
}$ and performing summation, $\sum_{k=1}^{N}\epsilon _{ik}[\dots ]_{k}$, we
obtain, upon using (\ref{45}), a general solution of the $N$-component
linearly degenerate hydrodynamic reduction in an implicit form (cf. (\ref%
{theo}))
\begin{equation}
x+\xi _{i}t=P_{i}(r^{i})-r^{i}P_{i}^{\prime }(r^{i})-\underset{m\neq i}{\sum
}\epsilon _{im}P_{m}^{\prime }(r^{m}),\text{ \ }i=1,2,...,N,  \label{multi}
\end{equation}%
where $P_{i}(r^{i})$, $i=1,\dots ,N$, are arbitrary functions.

Note that under the re-parametrization
\begin{equation*}
P_{k}^{\prime \prime }(\xi )=-\frac{\phi _{k}(\xi )}{f(\xi )}
\end{equation*}
the generalised hodograph solution (\ref{multi}) becomes
\begin{equation}  \label{fer1}
x+\xi _{i}t=\overset{r^{i}}{\int }\frac{\xi \phi _{i}(\xi )}{f(\xi )}d\xi +%
\underset{m\neq i}{\sum }\epsilon _{im}\overset{r^{m}}{\int }\frac{\phi
_{m}(\xi )}{f(\xi )}d\xi \, .
\end{equation}
Now, comparison of (\ref{fer1}) with the Ferapontov \cite{fer91} solution (%
\ref{theo}) provides a direct way for the identification of the entries of
the St\"{a}ckel matrix (\ref{stak}). Also, for this choice of the St\"{a}%
ckel matrix all constants $C_{l,m}$ (see Section 6) can be expressed in
terms of the coefficients $\epsilon _{ij}$ and $\xi _{k}$ by (\ref{theorem52}%
), (\ref{array}).

For the particular choice of $f(\xi )$ defined as
\begin{equation}
f(\xi )=\sqrt{R_{K}(\xi )}  \label{root}
\end{equation}%
where
\begin{equation*}
R_{K}(\xi )=\overset{K}{\underset{n=1}{\prod }}(\xi -E_{n})\,,
\end{equation*}%
and $E_{1}<E_{2}<\dots <E_{K}$ are real constants ($K=2N+1$ if $N$ is odd
and $K=2N+2$ if $N$ is even); and $\phi _{k}(\xi )$ being arbitrary
polynomials in $\xi $ of degrees less than $N$, system (\ref{multi})
describes quasiperiodic solutions of the form
\begin{equation}
x+\xi _{i}t=\overset{r^{i}}{\int }\frac{\xi \phi _{i}(\xi )d\xi }{\sqrt{%
R_{K}(\xi )}}+\underset{m\neq i}{\sum }\epsilon _{im}\overset{r^{m}}{\int }%
\frac{\phi _{m}(\xi )d\xi }{\sqrt{R_{K}(\xi )}},\text{ \ }i=1,2,...,N,
\label{quasi1}
\end{equation}%
The proof of quasiperiodicity of solution (\ref{quasi1}) is analogous to
that for solution (\ref{1}), (\ref{2}), (\ref{3}) obtained for $N=3$.

\subsection{Linearly degenerate commuting flows}

To extract the family of linearly degenerate commuting flows from general
representation (\ref{comchar}) we formulate the following

\textbf{Lemma 8.2}: \textit{For the linearly degenerate commuting flows each
function} $P_{i}(r^{i})$ \textit{in (\ref{comchar}) is linear with respect
to the corresponding Riemann invariant} $r^{i}$.

\textbf{Proof}: The condition $\partial _{i}W^{i}=0$ of linear degeneracy of
the commuting flow implies, on using (\ref{ld}) and (\ref{com}), that $%
P_{i}^{\prime \prime }(r^{i})=0$.

\medskip

We now consider the representation for the family of linearly degenerate
commuting flows suggested by the form of the kinetic equations (\ref{kin2}),
(\ref{s2}) for the KdV hierarchy. Importantly, the whole KdV kinetic
hierarchy (\ref{kin2}), (\ref{s2}) is characterised by a single integral
kernel, $G(\eta ,\mu )=\ln |(\eta -\mu )/(\eta +\mu )|$ (which is consistent
with the fact that all equations of the original finite-gap Whitham
hierarchy are associated with the same Riemann surface). This suggests that
there could exist a family of commuting flows to general nonlocal kinetic
equation (\ref{kin1}) having the form
\begin{equation}
\begin{split}
f_{\tau }& =(\tilde{s}f)_{x}\,, \\
\tilde{s}(\eta )& =\tilde{S}(\eta )+\frac{1}{\eta }\int\limits_{0}^{\infty
}G(\eta ,\mu )f(\mu )[\tilde{s}(\mu )-\tilde{s}(\eta )]d\mu \,,
\end{split}
\label{kincomm}
\end{equation}%
where $\tilde{S}(\eta )$ is an arbitrary function. Although verification of
commutativity of the kinetic equations (\ref{kin1}) and (\ref{kincomm}) is
beyond the scope of the present paper, it is clear that, if these equation
do commute, this must be manifested on the level of hydrodynamic reductions
as well. Having this in mind, we consider the $N$-component hydrodynamic
reductions to (\ref{kincomm}) obtained by the familiar delta-functional
ansatz (\ref{delta}) and try to see if they commute with the original
reductions (\ref{cont})--(\ref{rel1}).

First we notice that equation (\ref{kincomm}) is, essentially, the same
kinetic equation (\ref{kin1}) but with a different time variable and
different \textquotedblleft free soliton speed\textquotedblright\ function $%
S(\eta )$. Now, since we have proved integrability of the linearly
degenerate hydrodynamic reductions (\ref{cont})--(\ref{rel1}) in a general
form, we automatically have that analogous $N$-component hydrodynamic
reductions of (\ref{kincomm}) must also be integrable linearly degenerate
systems. It should be noted that, since the function $\tilde{S}(\eta )$ is
arbitrary, the set $\{\tilde{\xi}_{1},\dots ,\tilde{\xi}_{N}\}$ of its
values $\tilde{\xi}_{j}=\tilde{S}(\eta _{j})$ can be viewed as a set of $N$
arbitrary numbers, and the corresponding `cold-gas' hydrodynamic reduction
becomes (cf. (\ref{cont})--(\ref{rel1}))
\begin{equation}
u_{\tau }^{i}=(u^{i}\tilde{v}^{i})_{x},\qquad i=1,\dots ,N,  \label{cont1}
\end{equation}%
where the velocities $\tilde{v}^{i}=-\tilde{s}^{i}$ and the conservation law
densities $u^{i}$ satisfy algebraic relations
\begin{equation}
\tilde{v}^{i}=\tilde{\xi}_{i}+\sum_{k\neq i}^{N}\epsilon _{ik}u^{k}(\tilde{v}%
^{k}-\tilde{v}^{i})\text{, \ \ \ \ }\epsilon _{ik}=\epsilon _{ki}\text{, \ \
\ \ }\,  \label{alg1}
\end{equation}%
and $\epsilon _{ik}$ are the same as in (\ref{rel1}).

According to Theorem 3.1, system (\ref{cont1}), (\ref{alg1}) can be
represented in the Riemann form
\begin{equation}
r_{\tau }^{i}=\tilde{v}^{i}(\mathbf{r})r_{x}^{i}\text{, \ \ }i=1,2,...,N%
\text{; \ \ }  \label{rim1}
\end{equation}%
where the dependence $\tilde{v}^{i}(\mathbf{r})$ of the characteristic
velocities on the Riemann invariants is determined by the same formulae (\ref%
{main}) with the only difference that, one now replaces $\xi _{j}$ with $%
\tilde{\xi}_{j}$, i.e. we have
\begin{equation}
\tilde{v}^{i}=\frac{1}{u^{i}}\underset{m=1}{\overset{N}{\sum }}\tilde{\xi}%
_{m}\beta _{im}.  \label{vt1}
\end{equation}%
Indeed, representation (\ref{vt1}) is a straightforward consequence of (\ref%
{main}) since the rotation coefficients $\beta _{ij}$ and Lam\'{e}
coefficients $\bar{H}_{k}$ do not depend on the parameters $\xi _{m}$ (see (%
\ref{z}), (\ref{third}), the first formula in (\ref{main}), and
normalisation (\ref{norm})). It not difficult to see that commutativity
relationships (see (\ref{ts}))
\begin{equation}
\frac{\partial _{i}\tilde{v}^{j}}{\tilde{v}^{i}-\tilde{v}^{j}}=\frac{%
\partial _{i}v^{j}}{v^{i}-v^{j}}\,,\qquad i,j=1,2, \dots, N\, , \quad i\neq
j\,,  \label{ts3}
\end{equation}%
are satisfied identically. Thus, we have proved the following

\medskip

\textbf{Lemma 8.3}: \textit{Linearly degenerate semi-Hamiltonian flows (\ref%
{cont1}), (\ref{alg1}) and (\ref{cont}), (\ref{alg}) commute for any $N$}.

\medskip

In conclusion we note that, although we have proved integrability of the
`cold-gas' hydrodynamic reductions (\ref{cont}), (\ref{alg}) for an
arbitrary choice of the functions $S(\eta)$ and $G(\eta, \mu)$ in the
original kinetic equation (\ref{kin1}), one can expect that integrability of
the full equation (\ref{kin1}) would require some additional restrictions
imposed on the integral kernel $G(\eta, \mu)$ (other than just symmetry).

\section{Outlook and Perspectives}

Kinetic equation (\ref{kin1}) first arose as a continuum (thermodynamic)
limit of a semi-Hamiltonian hydrodynamic type system (the KdV-Whitham
system). This equation seems to belong to an entirely new
class of integrable systems, which we at present are unable to equip with
the standard attributes such as a Lax pair, commuting flows, Hamiltonian structures etc. This
paper makes the first step towards the understanding of the integrable
structure of equation (\ref{kin1}) by studying in detail the simplest class
of its hydrodynamic reductions and identifying them as {\bf the} Egorov,
semi-Hamiltonian linearly degenerate hydrodynamic type systems. The
availability of an infinite set of the aforementioned hydrodynamic
reductions is a strong evidence that the full equation (\ref{kin1})
could constitute an integrable system in the sense yet to be explored. While the studied `cold-gas' reductions
turn out to be integrable for an arbitrary symmetric `interaction kernel' $G(\eta ,\mu
)$,  integrability of the full equation (\ref{kin1}) will clearly require
some additional restrictions to be imposed on this kernel. Recent results \cite%
{ferkhus04a, ferkhus04b}, \cite{gibts96, gibts99} on the integrability of
2+1 hydrodynamic type systems and hydrodynamic chains, which are close
`relatives' of kinetic equations, suggest that these restrictions should be
determined by the condition of the existence, for an arbitrary $N$, of $N$%
-component hydrodynamic reductions parameterised by $N$ arbitrary functions
of a single variable. The most natural way to attack this problem is to
study the associated \textit{hydrodynamic chain}, i.e. an infinite set of the \textit{%
moment equations} (see e.g. \cite{gibrai07}) for
kinetic equation (\ref{kin1}). However, due to the structure of the nonlocal
term in (\ref{kin1}) the construction of this chain is far from being a
straightforward task.

The study of the moment equations for (\ref{kin1}) is also important in the
original context of the description of macroscopic dynamics of soliton
gases \cite{zakh71}, \cite{elkam05}. Indeed, the kinetic description of a soliton gas reflects the
particle-like nature of solitons. At the same time, one should remember that
solitons represent localized waves so the kinetic description of a soliton
gas should be complemented by the expressions for the averaged
characteristics of the underlying `microscopic' oscillatory wave field in
terms of the distribution function $f(\eta, x,t)$. Say, for the KdV equation
(\ref{kdveq}) the expressions for the two first moments of the wave field
have the form (see \cite{el03})
\begin{equation}  \label{mean}
\overline{\phi }(x,t)=4\int_{0}^{\infty }\eta f(\eta, x,t )d\eta \,,\quad
\overline{\phi^{2}}(x,t)=\dfrac{16}{3}\int_{0}^{\infty }\eta
^{3}f(\eta,x,t)d\eta \,
\end{equation}
and are identical to those arising in the Lax-Levermore-Venakides theory \cite%
{laxlev83}, \cite{ven87}, \cite{ven90} with the crucial difference that the dynamics of the distribution
function $f(\eta, x,t)$ is now governed by kinetic equation (\ref{kin1}%
), (\ref{kdv}) rather than the $N$-phase averaged Whitham equations (\ref{ffm}) so (\ref{mean}) are {\it ensemble averages}.

This paper was concerned mostly with the {\it structure} of the kinetic equation (\ref{kin1}). At the same time,
behaviour of its solutions and the associated evolution of the dynamical (moments, amplitudes etc.) and probabilistic (probability density, correlation function etc.) characteristics of the underlying rapidly oscillating wave field could be of considerable interest for applications. In this regard, we mention an interesting consideration following from our present study.
In the original construction \cite{el03} described in Section 3 the kinetic equation for the KdV soliton gas was obtained as the thermodynamic limit of the $N$-phase averaged KdV-Whitham equations (\ref{ffm}). These Whitham equations are {\it genuinely nonlinear} for any $N \in \mathbb{N} $ \cite{lev88}, i.e. for a reasonably general class of initial conditions the modulation dynamics specified on a Riemann surface of genus $N$ implies hydrodynamic breaking at some $t < \infty$ accompanied by
the growth of the genus $N$ (see e.g. \cite{DN}, \cite{ekv01}, \cite{gt02}). At the same time, the
`cold-gas' hydrodynamic reductions of the kinetic equations studied here are linearly degenerate, i.e. no breaking is expected and the number of gas components does not change during the evolution (a simple example of such non-breaking evolution for a two-component soliton gas was considered in \cite{elkam05}). Of course, there is no contradiction between these two contrasting types of behaviour as  the kinetic equation (\ref{kin11}), (\ref{s1}) represents a {\it singular} limit as $N \to \infty$ of the KdV-Whitham equations  while their genuine nonlinearity property is established only for finite $N$. Construction of physical solutions to linearly degenerate multi-component hydrodynamic type system (\ref{01}), (\ref{02}) and study of the associated wave field dynamics of  soliton gases in various integrable systems represents a separate interesting mathematical problem, which could find applications in the description of propagation and interaction of quasi-monochromatic soliton beams in dispersive dissipationless media.

 Another challenging problem is derivation of the 2+1 dimensional kinetic equation for the soliton gas in the framework of the Kadomtsev - Petviashvili (KP-2) equation. This problem would require computing the thermodynamic limit of the KP-Whitham equations associated with general algebraic (not necessarily hyperelliptic) Riemann surfaces \cite{krich1}, \cite{krich2}.

Finally, we would like to mention one more perspective arising from our study. To
our best knowledge, nonlocal kinetic equation (\ref{kin1}) is the first
available example of a continuum limit of a semi-Hamiltonian hydrodynamic
type system. The key point of its derivation is that it is not sufficient to
simply tend the number of Riemann invariants to infinity but it is important to
prescribe a special scaling controlling the distances between neighboring
invariants (in the case of averaged finite-gap dynamics, the widths of
spectral bands and gaps -- see (\ref{scal})). We believe that a similar
approach could be applied to a large class of semi-Hamiltonian hydrodynamic
type systems (not necessarily arising as the result of the Whitham averaging).
Of course, the corresponding thermodynamic scaling (an analogue of
distribution (\ref{scal})) could be different.

\section*{Acknowledgments}

We are grateful to V.E. Zakharov for his interest in this work and a number
of enlightening comments. We thank Yu. Fedorov, E. Ferapontov, O. Morozov,
A. Neishtadt, Z. Popowicz, S. Tsarev and A. Veselov for stimulating
discussions. The work has been partially supported by EPSRC (UK)(grant
EP/E040160/1) and London Mathematical Society (Scheme 4 Collaborative Visits
Grant). Work of M.V.P. has been also supported by the Programme
\textquotedblleft Fundamental problems of nonlinear
dynamics\textquotedblright\ of Presidium of RAS. M.V.P. and S.A.Z. also
acknowledge partial financial support from the Russian--Taiwanese grant
95WFE0300007 (RFBR grant 06-01-89507-HHC) .

\end{document}